\documentclass{aa} 
\usepackage{graphicx}

\newcommand\wise{{WISE~}}
\newcommand\wone{$W1$~}
\newcommand\wtwo{$W2$~}
\newcommand\wthree{$W3$~}
\newcommand\wfour{$W4$~}
\newcommand\wisep{WISE} 
\newcommand\wonep{$W1$}
\newcommand\wtwop{$W2$}
\newcommand\wthreep{$W3$}
\newcommand\wfourp{$W4$}
\newcommand\spitzer{{\it Spitzer~}}

\begin{document}

\title{Mid-infrared dust in two nearby radio galaxies, 
           \object{NGC 1316} (\object{Fornax A}) and \object{NGC 612} (\object{PKS 0131-36})
}

\author{B. Duah Asabere\inst{1}\inst{2}\inst{3}
\and C. Horellou\inst{2}
\and T. Jarrett\inst{4}
\and H. Winkler\inst{1}
}

\institute{Department of Physics, University of Johannesburg, P.O. Box 524, 
  Auckland Park 2006, Johannesburg, South Africa.
  \and Department of Earth and Space Sciences, 
  Chalmers University of Technology, 
  Onsala Space Observatory,
  SE-439 92 Onsala, Sweden  
\and
   Current address: Ghana Space Science and Technology Institute, 
   Ghana Atomic Energy Commission, P.O. Box LG 80, Legon, Accra, Ghana
\and
  Astronomy Department, University of Cape Town,  Private Bag X3,
  Rondebosch 7701, South Africa 
}

\date{}

\abstract 
{
Most radio galaxies are hosted by giant gas-poor ellipticals, but some contain significant amounts of dust, which is likely to be of external origin.
}
{
In order to characterize the mid-IR properties of two of the most nearby and brightest merger-remnant radio galaxies of the Southern hemisphere, NGC~1316 (Fornax~A) and NGC~612 (PKS~0131--36), 
we used observations with the Wide-field Infrared Survey Explorer (\wisep) 
at wavelengths of 3.4, 4.6, 12 and 22~$\mu$m  
and \spitzer mid-infrared spectra.  
}
{
By applying a resolution-enhancement technique, new \wise images were produced  
at angular resolutions ranging from   
$2\farcs6$ to $5\farcs5$. Global measurements were performed in the four \wise bands, and stellar masses and star-formation rates were estimated using published scaling relations.  
Two methods were used to uncover the distribution of dust, one relying on two-dimensional fits to the 3.4~$\mu$m images to model the starlight, and the other one using a simple scaling and subtraction of the 3.4~$\mu$m images to estimate the stellar continuum contribution to the emission in the 12 and 22~$\mu$m bands. 
} 
{
The two galaxies differ markedly in their mid-IR properties.  
The 3.4~$\mu$m brightness distribution can be well represented by the superposition of two S\'ersic models in NGC~1316 and by a S\'ersic model and an exponential disk in NGC~612.  
The \wise colors   
of NGC~1316 are typical of those of early-type galaxies; 
those of NGC~612 are in the range found for star-forming galaxies. 
From the 22~$\mu$m luminosity, we infer a star-formation rate of $\sim0.7~M_{\odot}$~yr$^{-1}$ in NGC~1316 and $\sim7~M_\odot$~yr$^{-1}$ 
in NGC~612. 
\spitzer spectroscopy shows that the 7.7-to-11.3~$\mu$m PAH line ratio is significantly lower in NGC~1316 than in NGC~612. 
The \wise images reveal resolved emission from dust in the central 1$'$--2$'$ of the galaxies. In NGC~1316, the extra-nuclear emission coincides with two dusty regions NW and SE of the nucleus seen in extinction in optical images and where molecular gas is known to reside. 
In NGC~612 it comes from a warped disk.  
This suggests a recent infall onto NGC~1316 and disruption of one or several smaller gas-rich galaxies, but a smoother accretion in NGC~612.  
While the nucleus of NGC~1316 is currently dormant and the galaxy is likely to evolve into a passive elliptical, 
NGC~612 has the potential of growing a larger disk and sustaining an active nucleus. 
}  
{
NGC~1316 and NGC~612 represent interesting 
challenges to models of formation and evolution of galaxies and AGNs.
}

\keywords{Galaxies: individual: NGC~1316 -- 
Galaxies: individual: NGC~612 -- 
Galaxies: ISM --
Galaxies: interactions -- 
Infrared: galaxies
}

\titlerunning{
Dust in NGC~1316 and NGC~612
} 

\authorrunning{Duah Asabere et al.} 

\maketitle

\section{Introduction} \label{sec:intro}

\begin{figure*}
\includegraphics[width=8.8cm]{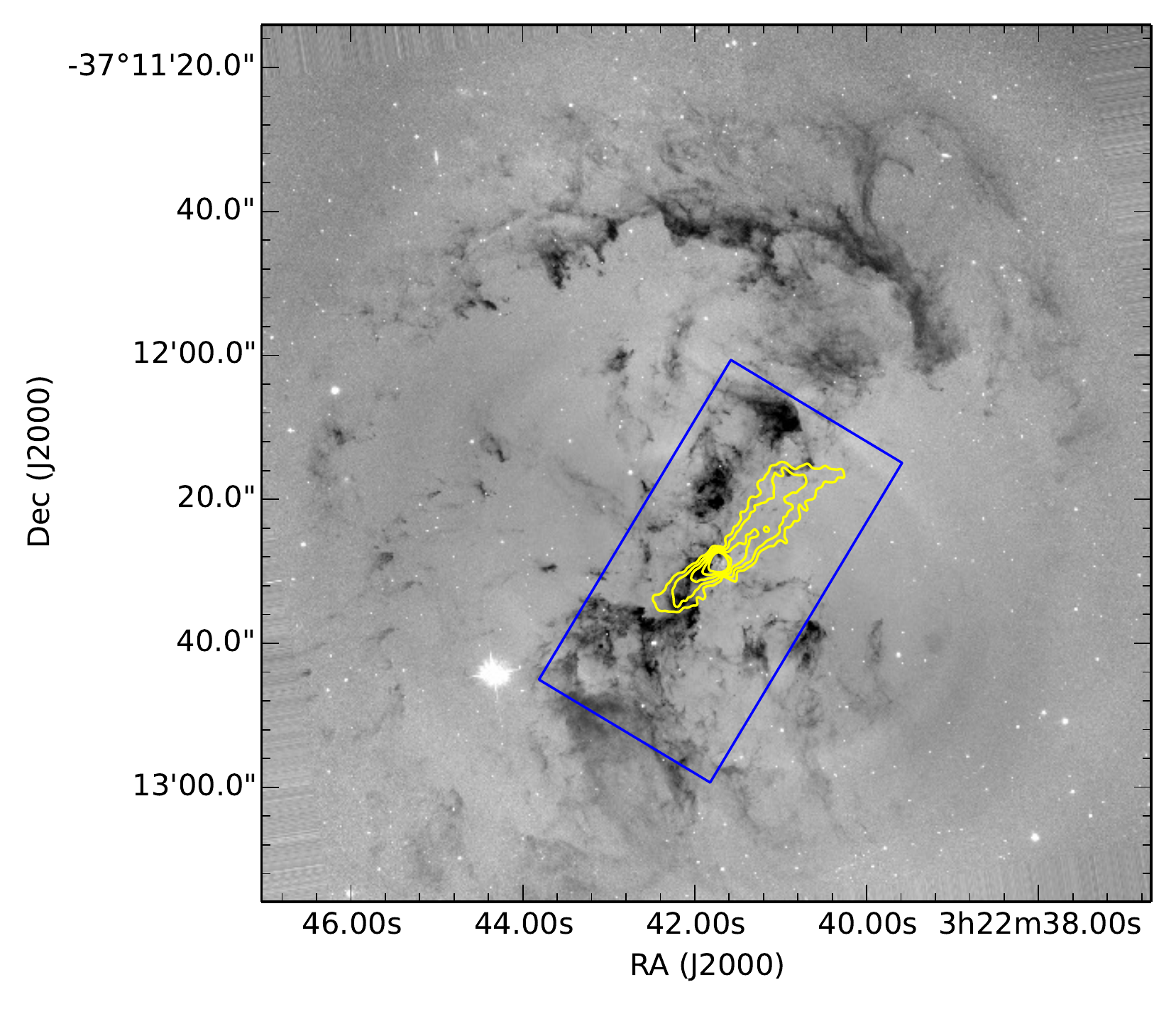}
\includegraphics[width=8.8cm]{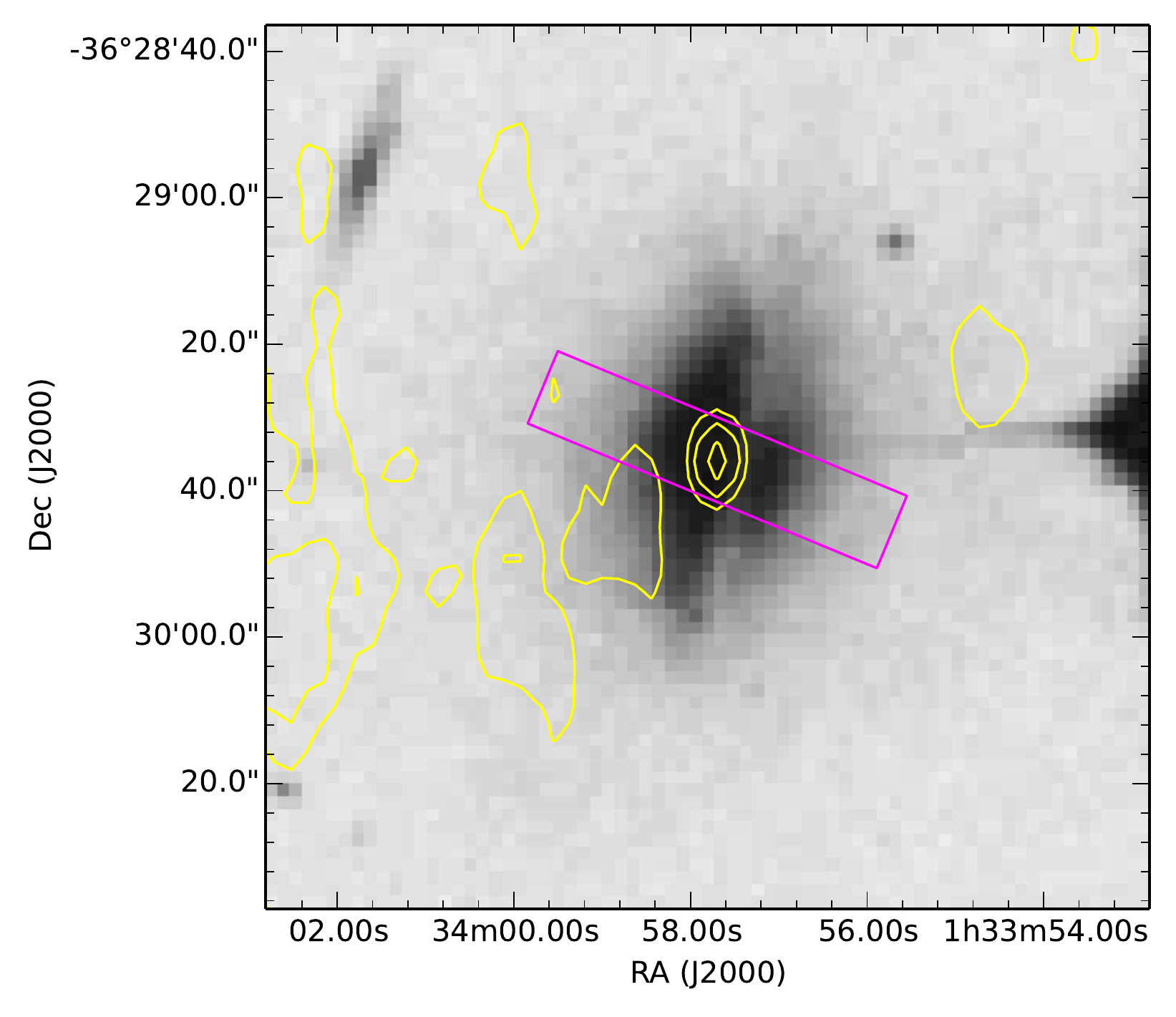}
\caption{
Optical images of the central region of NGC~1316 (left panel) and NGC~612 (right panel) in inverted gray scale and overlaid contours of the radio emission in yellow. The size of the images is $2'\times2'$ and the bright extended radio lobes are located outside the pictures. 
The rectangles show the regions from which the \spitzer IRS spectra displayed in Fig.~\ref{figIRSbothgals} were extracted. 
For NGC~1316, the size of the rectangle is $27.8''\times51.8''$ with a position angle (PA) of $149^\circ$. 
For NGC~612, it is $10.7''\times51.6''$ with a PA of 67.5$^\circ$. 
{\it Left panel:} HST image of NGC~1316 with 4.9~GHz radio contours of the inner jet overlaid \citep{1984AJ.....89.1650G}. 
The radio contours increase by a factor of 2 from 0.2 to 3.2~mJy/beam, where the beam is $1.38''\times1.02''$, with a PA of 20.3$^\circ$. 
{\it Right panel}: Digitized Sky Survey image of NGC~612. 
The 6~cm radio contours are 5, 10 and 20~mJy/beam and the beam is $7.8''\times4.6''$ with a PA of $-1.3^\circ$ 
(a larger radio continuum image of NGC~612 can be seen in Fig.~6 of \citealt{2008MNRAS.387..197E}). 
Note that NGC~612 is about six times more distant than NGC~1316.
}
\label{figHSTDSS}
\end{figure*}

Powerful radio galaxies are living evidence of relatively recent activity in the nuclei of their host galaxies. The characteristic timescale for ejected plasma to reach the outer radio lobes, located some hundreds of kpc away from the center, is a few tens to hundreds of Myr. Most observations are consistent with the scenario that radio lobes can be triggered during the merger of two gas-rich galaxies, a process that also results in the formation of a large elliptical galaxy with a supermassive black hole
(e.g. \citealt{1991ApJ...370L..65B}, 
\citealt{2012ApJ...758L..39T}). 
Most radio galaxies are hosted by gas-poor giant elliptical galaxies, but some contain significant amounts of gas and dust in their central regions; this material is likely of external origin, injected during the more recent infall of low-mass gas-rich galaxies onto the giant elliptical. 
At intermediate redshift ($0.05 < z < 0.7$), the median dust mass of bright radio galaxies  
is $1.2\times 10^9$~$M_\odot$, which is 
about twice as large as in the Large Magellanic Cloud 
but about four times lower than in the Milky Way 
\citep{2014MNRAS.445L..51T}. 
The spread in dust mass in that sample is however very large (more than two orders of magnitudes), which probably reflects the large range in the total mass and morphological type of the progenitors. 

Whereas mergers might be a necessary condition for an Active Galactic Nucleus (AGN) and the ejection of radio lobes, it is clearly not sufficient: only a fraction of elliptical galaxies with signs of merging activity and recent star formation are radio-loud \citep{2012MNRAS.419..687R}. 
The temperature of the gas fueling the AGN (either hot and associated with the surrounding X-ray gas, or cold and associated with atomic or molecular gas) probably plays a role in the level of activity of the nucleus 
\citep{2007MNRAS.376.1849H}. 

An interesting recent development is the increasing number of radio galaxies found associated with disk-like galaxies (e.g. \citealt{2015MNRAS.447.2468H},  
\citealt{2015MNRAS.454.1556S}). 
Those galaxies, although very rare, challenge the standard model of radio galaxy formation. Some of them (Centaurus A and Fornax A) are giant ellipticals with dust lanes oriented along the minor axis; others have a clear disk seen edge-on 
(e.g. NGC~612, \citealt{1978A&A....69L..21E}; 
2MASX~J03155208-1906442, \citealt{2001ApJ...552..120L}; 
Speca, \citealt{2011MNRAS.417L..36H}; 
J2345-0449, \citealt{2014ApJ...788..174B}; 
NGC~1534, \citealt{2015MNRAS.447.2468H}). 
It is likely that the latter have had a quieter secular evolution, possibly via slow accretion rather than  mergers, to allow the disk to grow and gain angular momentum at the same time as the central black hole increased its mass. 
Simulations indicate that the angular momentum of the gas when it enters a halo is an important factor for the subsequent accretion by the black hole 
\citep{2013ApJ...779..136B}. 

Clearly, the central regions of the hosts of radio galaxies contain valuable information about their recent history. Mid-IR imaging provides additional information on the AGN, stars, and interstellar medium and permits investigation of their inter-relation. Data taken by the Wide-field Infrared Survey Explorer 
(\wisep; \citealt{2010AJ....140.1868W}) 
are well suited for that purpose. The short wavelength \wise bands,  \wone and \wtwop, are dominated by emission from an evolved stellar population, while the \wthree and \wfour bands primarily capture emission from warm dust ($T > 30$~K), potentially at least partially heated by an AGN 
\citep[see][]{2011ApJ...735..112J}.
Polycyclic aromatic hydrocarbon (PAHs) molecular features at 
11.3 and 12.7~$\mu$m and nebular [NeII] line emission at 12.8~$\mu$m may also contribute to the flux in the \wthree band.  
At low redshifts, the \wfour band is mostly free from emission lines and traces warm dust and possible AGN. 

In this paper, we analyze \wise mid-IR images of 
NGC~1316 (Fornax~A) and NGC~612, two prominent nearby examples of dusty radio galaxies in the Southern hemisphere.
Whereas the two galaxies are rather similar in their radio properties, with bright radio lobes extending out to large distances from the galaxies' centers 
(about 150~kpc on the sky for NGC~1316 and 250~kpc for NGC~612), 
they are very different in the optical 
(see Fig.~\ref{figHSTDSS}). 
 NGC~612, of morphological type  SA0$^+$~pec \citep{1991rc3..book.....D}, 
possesses a dusty stellar disk with a young stellar population 
\citep{2007MNRAS.381..611H}, 
which is very unusual for a radio galaxy
(\citealt{1978A&A....69L..21E}, \citealt{2001A&A...375..791V}). 
It also possesses an enormous 
H{\textrm{I}} disk   
($\sim140$~kpc wide, or $\sim4'$; 
$1.8\times10^9 M_\odot$, \citealt{2008MNRAS.387..197E}). 
In contrast, NGC~1316, of morphological type SAB(s)0$^0$~pec \citep{1991rc3..book.....D}, 
 is a giant galaxy with outer shells and loops and an irregular dust pattern in the central 2$'$ 
(e.g. \citealt{1980ApJ...237..303S}, \citeyear{1981ApJ...246..722S}), 
and it is poor in atomic gas ($\sim 2\times 10^7 M_\odot$,  
\citealt{2001A&A...376..837H}). 
Both galaxies contain molecular gas: 
about $6\times 10^{8} M_\odot$ 
in  NGC~1316 
(\citealt{2001A&A...376..837H}, 
after correction for our adopted distance; 
see also \citealt{1993ApJ...419..544S}) 
and $7.9\times10^9 M_\odot$ in NGC~612 
\citep{2012JPhCS.372a2067P}.  
NGC~612 has the highest H$_2$ mass in the 
\citealt{2012JPhCS.372a2067P} 
sample\footnote{
The \cite{2012JPhCS.372a2067P} 
sample  
 was extracted from the flux-density--limited sample of radio galaxies 
of \citet{1989MNRAS.236..737E} from the Parkes 2.7~GHz survey in the declination range --40$^\circ$ to --17$^\circ$. It includes NGC~612 and NGC~1316.} 
of 11 nearby ($z < 0.03$) southern early-type radio galaxies. 
A brief description of the galaxies is given in Sect.~\ref{sec:thegalaxies}. 

Preliminary results 
based on lower-resolution \wise images of NGC~1316 
were reported by \cite{2014arXiv1409.2474D}. 
The data presented here have been reprocessed in order to achieve a higher angular resolution, 
as described in Sect.~{\ref{sec:data-analysis}}. 
In Sect.~{\ref{sec:flux}} the results 
from our global flux measurements in the four \wise bands are presented. 
In Sect.~\ref{sectSpitzerIRS} archival \spitzer IRS spectra of the central regions of the galaxies are shown and discussed in relation to the \wise observations. 
Sect.~{\ref{sec:composep}} is a description of the 
models of the galaxies' starlight that were tested 
and of the two different methods that were used to separate the contributions of the various components (starlight, AGN, dust) 
to the mid-IR emission. 
Since NGC~1316 has been observed with \spitzer at 3.6, 4.5, 5.8, 8 and 24~$\mu$m 
(\citealt{2005ApJ...622..235T}, \citealt{2010ApJ...721.1702L}), 
we focus here on the WISE \wthree 12~$\mu$m image. 
In Sect.~\ref{sec:SMASS-SFR} the \wise colors  
and estimates of the stellar masses and star-formation rates are given.  
Finally, the results and the methods are discussed in Sect.~\ref{secDiscussion}. 

All magnitudes are quoted in the Vega system.

\begin{figure*}
\includegraphics[height=8.0cm]{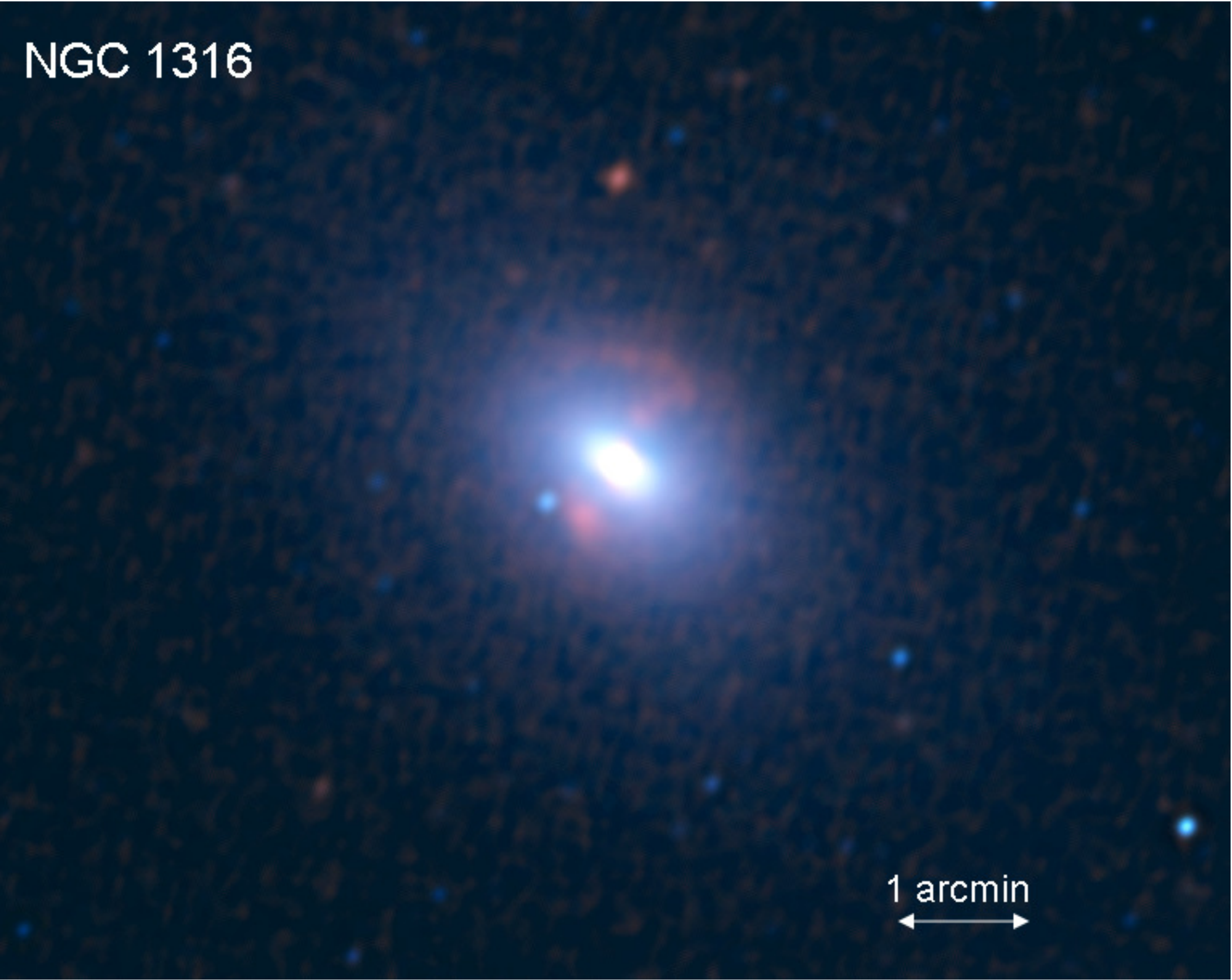} 
\includegraphics[height=8.0cm]{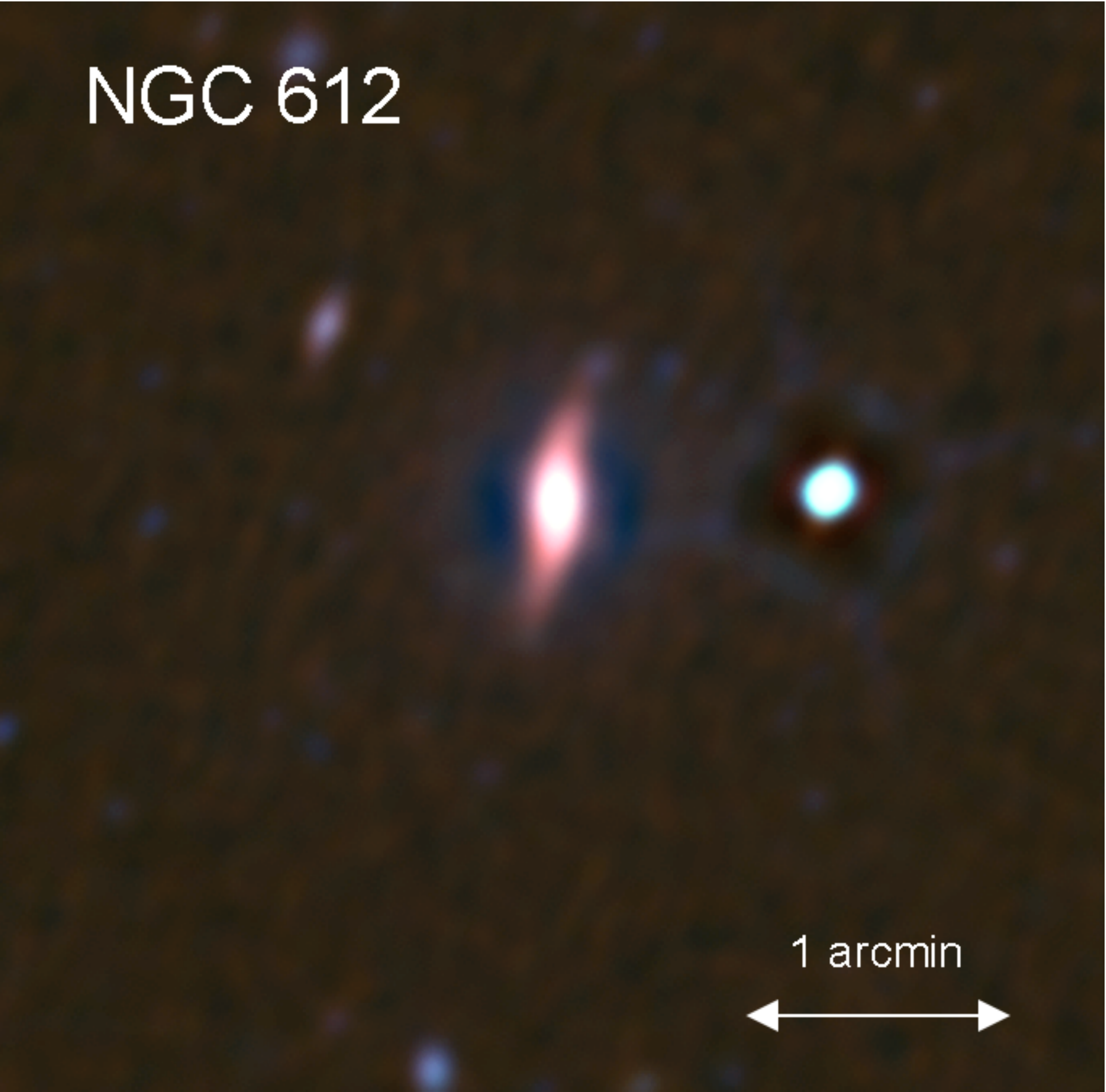}
\caption{High-resolution three-color WISE images (\wone in blue, \wtwo in green, \wthree in red) of NGC~1316 (left panel) and NGC~612 (right panel). 
The dust lanes are clearly seen. 
The super-resolution technique creates artifact features around high surface brightness emitters (e.g. nuclei, stars; see Sect.~\ref{sec:data-analysis} for further details). 
1 arcminute corresponds to 5.8~kpc in NGC~1316 and to 36.6~kpc in NGC~612.
}
\label{fig:hires-wise} 
\end{figure*}

\begin{table*}[t]
 \centering
 \caption{Some characteristics of the enhanced-resolution (MCM-HiRes) \wise maps.} 
 \begin{tabular}[t]{c c c c c}
 \hline \hline
Band\tablefootmark{a} 
& Wavelength\tablefootmark{b} 	
& Resolution\tablefootmark{c}
& Sensitivity\tablefootmark{d}
& Sensitivity\tablefootmark{e}\\
&($\mu$m)& (arcsec) & (mag~arcsec$^{-2}$) &($\mu$Jy~arcsec$^{-2}$)\\
 \hline 
 \wone & 3.4 	& 2.6	&23.4  &0.14\\
 \wtwo & 4.6 	& 3.0	&22.4  &0.19\\
 \wthree& 12.0 	& 3.5 	&18.6  &1.1\\
 \wfour & 22.8 	& 5.5	&16.1  &2.9\\   
 \hline
 \end{tabular}
 \tablefoot{
\tablefoottext{a}{\wise bands.}\\
\tablefoottext{b}{Central wavelengths \citep[taken from][]{2014PASA...31...49B}.} \\ 
\tablefoottext{c}{Angular resolution.
For comparison, the original single frame \wise images have a diffraction and optics-limited resolution 
of $\sim 6''$ in the three short-wavelength bands and $\sim 12''$ in the 22~$\mu$m band  
 \citep[see][]{2012AJ....144...68J}. 
The enhanced resolution of the \wise images is comparable to that of \spitzer images 
($\sim 2.5''$ for IRAC and $\sim 6''$ for MIPS-24). 
}\\
\tablefoottext{d}
{Approximate $1\sigma$ surface brightnesses 
(Vega) measured in the NGC~1316 mosaic images.}\\
\tablefoottext{e} 
{As in (d), but in $\mu$Jy~arcsec$^{-2}$.
The pixel size is 0.6875$''$ in all four bands. }
}
 \label{tab:WISE-bands}
\end{table*}
\section{The galaxies} \label{sec:thegalaxies}

At a distance\footnote{
Several slightly conflicting distance estimates to NGC~1316 have been published 
(e.g. 
\citealt{2007ApJ...657...76F}, 
\citealt{2010AJ....140.2036S}, 
\citealt{2013A&A...552A.106C}). 
For simplicity, we adopt a distance of 20~Mpc for NGC~1316, 
consistent with the various estimates given in the NASA Extragalactic Database (NED).
At that distance, $10''$ corresponds to 0.97~kpc. 
} 
of about 20~Mpc, NGC~1316 (Fornax~A, PKS~0320--37, Arp~154) has been abundantly studied. 
The spectacular dust lanes of NGC~1316 were discussed by Schweizer  
(\citeyear{1980ApJ...237..303S}, \citeyear{1981ApJ...246..722S}) 
and their detailed structure was revealed in {\it Hubble} Space Telescope (HST) images 
(\citealt{1999AJ....117..167G}; 
see also \citealt{2010Ap&SS.327..267C}). 
Some of the molecular gas was found about $40''$ NW of the center 
in the region of the dust lane 
where the central radio jet 
abruptly stops and may have been deviated 
(\citealt{2001A&A...376..837H},  
\citealt{1984AJ.....89.1650G}). 

A remarkable correspondence was found between 
the HST extinction features and 
the dust lanes seen in emission 
in the 
the Infrared Space Observatory (ISO) 15~$\mu$m image 
\citep{2004A&A...416...41X} 
and in the \spitzer 8~$\mu$m image \citep{2005ApJ...622..235T}. 
In a detailed multi-component analysis of near- and mid-IR \spitzer images, {\it Chandra} and XMM-{\it Newton} X-ray images, millimeter and radio maps, 
\cite{2010ApJ...721.1702L} 
quantified the inter-relations between the central AGN, the inner jet, the surrounding hot gas, and the dust and cool gas 
 likely due to the infall of a gas-rich companion less than $\sim 0.5$~Gyr ago  
(\citealt{1998AJ....115..514M}). 
From the analysis of the globular cluster population in NGC~1316, 
\cite{2004ApJ...613L.121G} 
estimated that a major merger took place about 3~Gyr ago. 
NGC~1316 is a member of the SINGS and KINGFISH samples\footnote{
SINGS (\spitzer Infrared Nearby Galaxies Survey) and KINGFISH (Key Insights on Nearby Galaxies: A Far-Infrared Survey with
Herschel) are samples of 75 and 61 nearby star-forming galaxies that span a wide range of luminosities, optical-to-infrared ratios and morphologies observed with \spitzer and {\it Herschel} 
(Kennicutt et al. \citeyear{2003AAS...203.9010K}, \citeyear{2011PASP..123.1347K}).  
}. 
{\it Herschel} global flux measurements between 70 and 500~$\mu$m were reported by \cite{2012ApJ...745...95D}, 
and resolved {\it Herschel} maps and APEX-LABOCA 870~$\mu$m observations were presented by Galametz et al. 
(\citeyear{2012MNRAS.425..763G}, 
 \citeyear{2014MNRAS.439.2542G}). 
By modeling the spectral energy distribution (SED), \cite{2007ApJ...663..866D} 
estimated the dust mass of NGC~1316 to be about 2.5~$\times 10^7 M_\odot$. 

As mentioned in the introduction, NGC~612 (PKS~0131--36) possesses a large HI disk and is about ten times richer than NGC~1316 in molecular gas. 
At 125~Mpc, it is about six times more distant than NGC~1316.
At that distance, 10$''$ correspond to 6.1~kpc.  
NGC~612 was detected in the four IRAS bands 
(\citealt{1989ApJS...70..329K}; see also corrections in NED by Knapp 1994), 
by ISO at 170~$\mu$m 
(\citealt{2004A&A...422...39S}) 
and by Akari (\citealt{2007PASJ...59S.401O}) at 9~$\mu$m and 18~$\mu$m \citep{2012ApJ...754...45I}. 
NGC~612 can be classified as a luminous infrared galaxy (LIRG) with a luminosity of 
$L_{8-1000~\mu{\rm m}} = 1.2\times 10^{11}\, L_\odot$, 
whereas NGC~1316 is about 25 times less luminous, $4.9\times 10^9\, L_\odot$   
(from the relation with the IRAS fluxes given by \citealt{1996ARA&A..34..749S} 
and using the corrected IRAS values in NED). 

\section{Data analysis} \label{sec:data-analysis}  


\begin{table*}[t]
\centering
\caption{WISE one-sigma isophotal-aperture photometry of NGC~1316.} 
\begin{tabular}[t]{l c c c c c c c c c c}
\hline \hline
Band  
& Flux density\tablefootmark{a} 
& Magnitude\tablefootmark{b} 
& $R_{\rm iso}$\tablefootmark{c}
& $R_{\rm eff}$\tablefootmark{d}
& $SB_{\rm eff}$\tablefootmark{e} 
& $C$\tablefootmark{f} 
& $SB_{\rm central}$\tablefootmark{g} 
& $PA$\tablefootmark{h}
& $AR$\tablefootmark{i}\\
& (mJy)& (mag) &(arcsec) & (arcsec) & (mag~arcsec$^{-2}$)& &(mag~arcsec$^{-2}$)& ($^\circ$)\\
\hline
    \wone  & ${2843.10\pm29.82}$ & $5.08\pm0.01$ & 761.1 & 101.3 &  
    16.61 & 7.66	& 13.58 & 34.7 &0.65\\
    \wtwo & ${1488.78\pm15.62}$ & $5.15\pm0.01$ & 588.7 &  93.4 & 
    16.50 & 7.05	& 13.61	& 34.7& 0.65\\
    \wthree  & ${449.90\pm4.78}$ & $4.53\pm0.01$ & 188.0 & 62.2 &  
    14.82 & 5.54	& 12.67	& 34.7	& 0.65\\
    \wfour  & ${341.84\pm4.27}$ & $3.46\pm0.01$ & 101.6 &  39.5 & 
    12.73 & 4.79	& 11.98	& 34.7	& 0.65\\
\hline
\end{tabular}
\tablefoot{
    \tablefoottext{a}{Integrated flux density.
Note that the quoted uncertainties do not include the uncertainty in the absolute calibration, 
which ranges from 2.4-5.7\% depending on the band 
(see section 4.4h/vii of the WISE All Sky Explanatory Supplement, \citealt{2012wise.rept....1C}).}\\ 
    \tablefoottext{b}{Integrated Vega magnitude.}\\
    \tablefoottext{c}{1-$\sigma$ isophotal radius or semi-major axis for the aperture photometry.}\\
    \tablefoottext{d}{Effective (half-light) radius.}\\
    \tablefoottext{e}{Effective (half-light) surface brightness.}\\
    \tablefoottext{f}{Concentration index (75\% to 25\% light ratio).}\\
    \tablefoottext{g}{Central surface brightness.}\\
    \tablefoottext{h}{and} 
    \tablefoottext{i}{are the position angle 
    and axis ratio of the \wone 3-$\sigma$ isophote.}\\
}
\label{tab:ngc1316-sources}
\end{table*}

\begin{table*}[t]
\centering
\caption{WISE one-sigma isophotal-aperture photometry of NGC~612.} 
\begin{tabular}[t]{l c c c c c c c c c c}
\hline \hline
Band
& Flux density\tablefootmark{a} 
& Magnitude\tablefootmark{b} 
& $R_{\rm iso}$\tablefootmark{c}
& $R_{\rm eff}$\tablefootmark{d}
& $SB_{\rm eff}$\tablefootmark{e} 
& $C$\tablefootmark{f} 
& $SB_{\rm central}$\tablefootmark{g} 
& $PA$\tablefootmark{h}
& $AR$\tablefootmark{i}\\
& (mJy)& (mag) &(arcsec) & (arcsec) & (mag~arcsec$^{-2}$)& &(mag~arcsec$^{-2}$)& ($^\circ$)&\\
\hline
    \wone & ${67.93\pm0.73}$ & $9.14\pm0.01$ & 141.3 & 16.7 &  16.65&   
    5.29 	& 20.57	&170.9 &0.60\\
    \wtwo  & ${39.64\pm0.43}$ & $9.09\pm0.01$ & 92.6 & 13.8 &  16.19 &    
    4.20	& 19.60	&170.9 &0.60\\
    \wthree  & ${113.90\pm1.23}$ & $6.02\pm0.01$ & 48.6 & 14.0 &  13.04 & 
    3.24	& 15.13	&170.9 &0.60\\
    \wfour  & ${144.52\pm2.19}$ & $4.40\pm0.02$ & 44.3 & 17.6 &  11.97 &   
    3.06	& 13.51	&170.9 &0.60\\
\hline
\end{tabular}
\tablefoot{As in Table~\ref{tab:ngc612-sources}.}
\label{tab:ngc612-sources}
\end{table*}

Mid-IR level 1-b frames of NGC~1316 and NGC~612 were downloaded from the \wise archive\footnote{
The All-sky Data Products and Atlas Images archive is maintained by the NASA IPAC (Infrared Processing and Analysis Centre) in the IRSA (Infra-Red Science Archive), 
{\tt{http://irsa.ipac.caltech.edu/applications/wise/}}
}. 
In order to improve the resolution of those images, we used   
the Image Co-addition with Optional Resolution Enhancement software package  
(ICORE, \citealt{2013ascl.soft02010M}) and a deconvolution technique based on 
the Maximum Correlation Method (MCM-HiRes, \citealt{2009ASPC..411...67M, 2012AJ....144...68J}). This allowed an improvement of a factor of 3--4 in angular resolution compared to the publicly released mosaics, which were optimized for point-source detection and have a degraded 
resolution of 30 to 40\% compared to the native single frames \citep[see][]{2012wise.rept....1C, 2013AJ....145....6J}. 
As shown in 
Table~\ref{tab:WISE-bands}, 
the achieved resolutions are about 3$''$ in the first three bands and 6$''$ in \wfourp, which is 
comparable (within about 30\%) to those of \spitzer IRAC and MIPS-24 images. 
For comparison, the resolution of the original \wise images is $\sim6''$ in the first three bands and $\sim 12''$ in the 22~$\mu$m band. 
This super-resolution reconstruction is not without limitations. Most commonly associated with deconvolution processing are ``ringing" (sinc function) artifacts associated with abrupt transitions between background (or low surface brightness emission) and bright emission such as stars or active galactic nuclei. These artifacts are mitigated to a certain extent (as described in \citealt{2012AJ....144...68J}), 
but are still present adjacent to the bright nucleus of NGC~1316 and the PSF-dominated inner regions of NGC~612.

The images were converted from Data Number (DN) into mJy/pixel by multiplying by the following factors, 
$1.9350\times 10^{-3}$, 
$2.7048\times 10^{-3}$, 
$1.8326\times 10^{-3}$,  and 
$5.2269\times 10^{-2}$  
for the respective \wonep, \wtwop,\wthreep, \wfour bands 
(\citealt{2010AJ....140.1868W}, 
\citealt{2012wise.rept....1C}). 

A color correction was applied to account for the steeply rising 
mid-IR spectra of dusty sources such as NGC~1316 and NGC~612. 
Rather than using the standard conversion for sources with a flat spectrum ($F_\nu \sim \nu^0$), we used  
the following values for the zero-magnitude flux density, $F_0$: 
$306.681$, 
$170.663$, 
$29.0448$, and 
$7.87$~Jy in the respective \wonep, \wtwop, \wthreep, \wfour bands  
(given by \citealt{2013AJ....145....6J} 
for the three first bands and 
by Brown et al. 2014 for \wfourp; the latter value takes into account the 
revised response function and effective wavelength of \wfourp). 


Figure~\ref{fig:hires-wise} shows three-color (\wonep, \wtwop, \wthreep) resolution-enhanced \wise images. 
The dust lanes are clearly visible in both galaxies. 

In order to characterize the various contributions to the mid-IR emission of the two galaxies, we used GALFIT, a two-dimensional image decomposition program that fits parametric functions to digitized images using standard chi-squared minimization \citep{2002AJ....124..266P}. 
As inputs, GALFIT uses, in addition to the images in the different bands, 
uncertainty maps and point-spread function (PSF) images to account for image smearing 
in the corresponding bands. 

The analysis was done on $6'\times6'$ images of NGC~1316 
and $4'\times4'$ images of NGC~612 centred on the galaxies' respective nuclei. 
The sky background emission in our enhanced-resolution and cleaned images, which virtually have no visible bad pixels to mask, was first subtracted. 
It was determined in an annulus well outside the target galaxy. 

\section{Global measurements} \label{sec:flux}

Source characterization of the \wise data was carried out using software tools specifically developed for \spitzer and \wise imaging, described in \citet{2013AJ....145....6J}. The most important steps included foreground star removal (PSF-subtraction), local background estimation using an elliptical-shaped annulus, ellipticity and position angle estimation from the 3$\sigma$ isophote (thereafter fixed for all radii), and the radial size based on the 1$\sigma$ isophote. Using a double S\'{e}rsic model \citep{1968adga.book.....S} to fit the galaxies' inner and outer regions, 
the total integrated flux was then computed from the integrated isophotal flux and the extrapolated light that extends beyond the 1$\sigma$ isophote. 
The flux density uncertainties reported in Tables~\ref{tab:ngc1316-sources} and \ref{tab:ngc612-sources} 
were computed from the background and Poisson error estimations. 
Finally, these global measurements were complemented with curve of growth and azimuthally-averaged surface brightness measurements that provide insights into internal structural changes with radius. 

Tables~\ref{tab:ngc1316-sources} and~\ref{tab:ngc612-sources} present the results. 
The flux density measurements of NGC~1316 compare very well with the corresponding \spitzer values in Table~1 of \citet{2005ApJ...633..857D}. 
For NGC~612, our measurement in \wfour is in agreement with the IRAS value at 25~$\mu$m 
($130\pm33$~mJy; NED); 
the \wthree isophotal flux is significantly lower than the 12~$\mu$m IRAS flux 
($200\pm35$~mJy; NED) and the 9~$\mu$m Akari flux of 135~mJy 
(\citealt{2012ApJ...754...45I}). 

For both galaxies, there is approximately a 50\% drop in the measured flux density from \wone to \wtwop; 
this is mainly due to the decline in stellar emission as one moves from 3.4~$\mu$m to 4.6~$\mu$m. 
NGC~1316 has a higher \wthreep-to-\wfour flux density ratio than NGC~612. 
This will be discussed in Sect.~\ref{secDiscussion}. 

\section{Spitzer IRS spectra of the central regions}
\label{sectSpitzerIRS}

\begin{figure*}
\includegraphics[width=18.0cm]{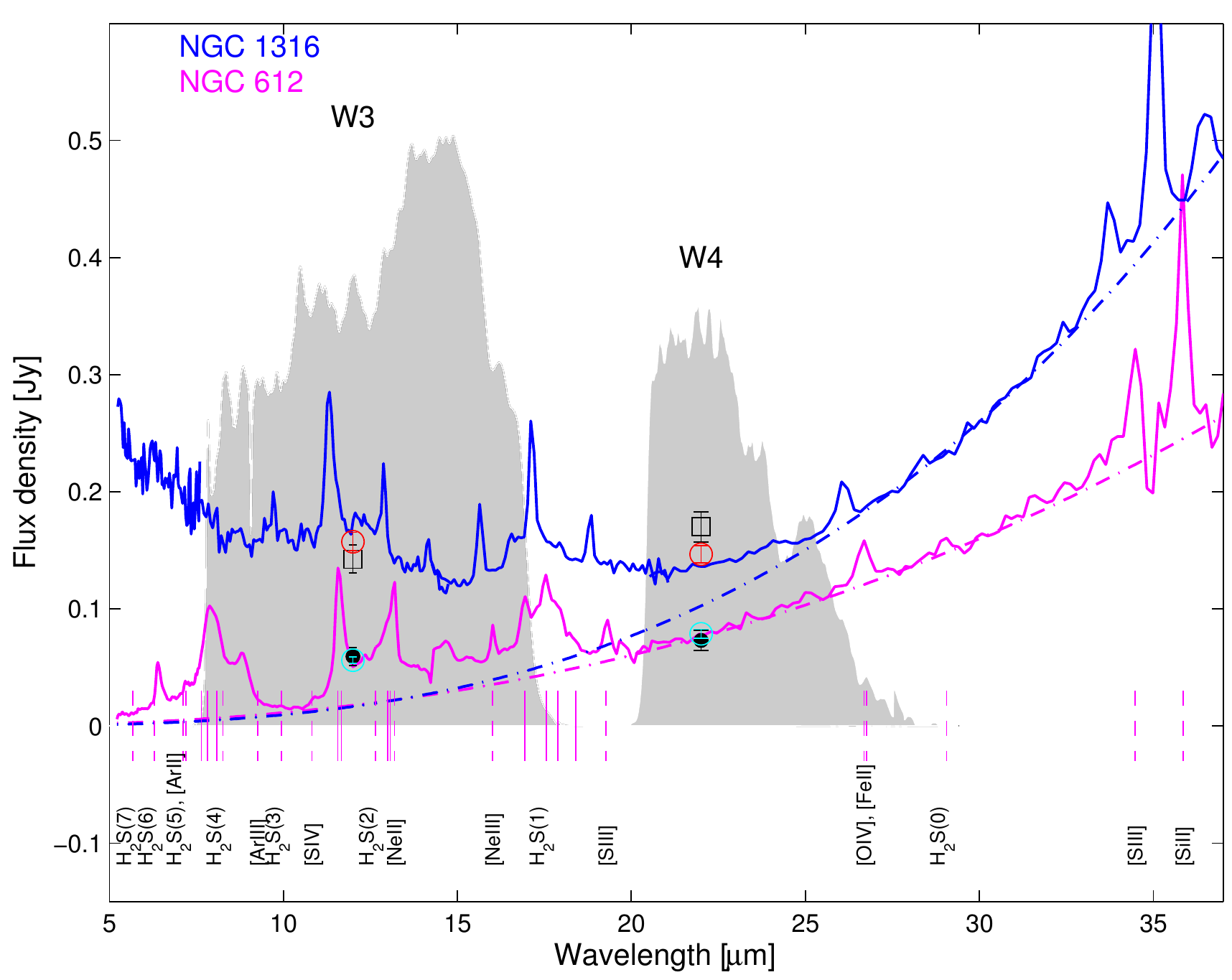}
\caption{\spitzer IRS spectra of NGC~1316 (in blue) and NGC~612 (in magenta). 
The spectra differ strongly at short wavelengths, where the Rayleigh-Jeans part of the stellar light is clearly seen for NGC~1316 whereas NGC~612 shows strong emission lines, including from PAHs at 7.7~$\mu$m.  
They also differ at long wavelengths, the NGC~1316 spectrum being significantly steeper than 
that of NGC~612, as illustrated by 
the dashed-dotted and the dashed lines which are power-laws with an index of $+3$ for NGC~1316 
and $+2.4$ for NGC~612. Note that those are not formal fits to the data. 
The response functions of the WISE \wthree and \wfour bands are shown in grey, in arbitrary units, 
taking into account the correction by \cite{2014PASA...31...49B}. 
The WISE fluxes measured in the rectangular regions used to extract the spectra are shown as squares for NGC~1316 and dots for NGC~612. 
The values obtained by weighing and summing up the \spitzer spectra over the WISE response functions are shown as open circles (red for NGC~1316 and cyan for NGC~612). 
The vertical lines indicate the wavelengths of the main lines (dashed) and of the PAHs (solid lines) at the redshift of NGC~612. 
}
\label{figIRSbothgals}
\end{figure*}

\spitzer IRS spectra are available for the central regions of both galaxies. 
The regions from which they were extracted are shown in 
Fig.~\ref{figHSTDSS} as rectangles overlaid on an HST image of NGC~1316 and on a Digitized Sky Survey image of NGC~612. 
Those regions are rather large and cover both the nuclei of the galaxies and 
areas where dust is seen in extinction. 
For NGC~1316, the extraction box is roughly oriented along the central dust lane, while in NGC~612 it is almost perpendicular to the dust lane of the nearly edge-on disk. The width of 
the extraction box of NGC~612 is $10\farcs7$ (6.5~kpc); in NGC~1316, the height of the box of $51\farcs8$ corresponds to 5~kpc.  

The spectra are a combination of four low-resolution spectra (SL2, SL1, LL2 and LL1, 
with increasing wavelength) and are shown 
in Fig.~\ref{figIRSbothgals} 
in the observer's frame. 
The NGC~1316 spectra were taken by the SINGS team \citep{2003PASP..115..928K} 
and downloaded from the SINGS archive\footnote{\tt irsa.ipac.caltech.edu/data/SPITZER/SINGS/galaxies}.
The NGC~612 spectrum was downloaded from the \spitzer archive as an Enhanced Imaging Product\footnote{
Project ID 30745, 
`Spitzer Observations of the First Unbiased AGN Sample of the Local Universe', 
PI K.I. Weaver}.  

The spectrum of NGC~1316 is shown here as a comparison to the NGC~612 spectrum and to help interpret the \wise broad-band observations. 
To our knowledge, the NGC~612 spectrum has not been published before. 
The near-IR to mid-IR spectrum of NGC~1316 has been discussed before 
(the \spitzer IRS spectrum by 
\citealt{2007ApJ...656..770S} 
and the 2.5--13~$\mu$m AKARI spectrum by 
\citealt{2007ApJ...666L..21K}).  

The response functions of the \wise \wthree and \wfour bands are displayed 
as shaded areas in Fig.~\ref{figIRSbothgals}, and the corresponding fluxes measured in those bands are shown as squares for NGC~1316 and black dots for NGC~612. 
We also calculated the flux contained in the IRS regions by weighing and summing up the \spitzer spectra over the response functions of the \wthree and \wfour bands. The resulting points are added to Fig.~\ref{figIRSbothgals} as open circles (red for NGC 1316 and cyan for NGC 612). 
The WISE broad-band measurements are in good agreement\footnote{
For NGC~612 we measured 
$56.2\pm2.8$~mJy in \wthree and 
$78.7\pm3.9$~mJy in \wfour from the \spitzer spectra. 
For NGC~1316 we got 
$157.4\pm7.9$~mJy and 
$146.6\pm 7.3$~mJy in the respective bands. 
The 5\% uncertainties reflect the uncertainties attached to the spectral values. 
For comparison, the values measured in the WISE maps are 
$59.4\pm7.7$~mJy and 
$73.0\pm8.5$~mJy for NGC~612, and 
$142.5\pm11.9$~mJy and 
$170.1\pm13.0$~mJy for NGC~1316.  
} 
with the \spitzer IRS spectra, especially for NGC~612. 

In NGC~1316, the low-wavelength part of the spectrum is dominated by the Rayleigh-Jeans part of the spectrum of the old stellar population, whereas NGC~612 displays strong emission lines, in particular the 7.7~$\mu$m PAH feature that enters the \wthree filter. 
\cite{2007ApJ...656..770S} 
had pointed out the peculiar spectrum of NGC~1316, with an unusually low ratio of the 7.7 to 11.3~$\mu$m line strength. 
The 7.7~$\mu$m PAH feature is produced by smaller, ionized PAHs whereas the 11.3~$\mu$m feature is due to larger and neutral PAHs (e.g. \citealt{1999ApJ...511L.115A}). 
Such low ratios were found in the other low-luminosity AGNs of the SINGS sample 
\citep{2007ApJ...656..770S} 
and in a sample of dusty elliptical galaxies 
\citep{2008ApJ...684..270K}. 
It is likely that the PAHs in NGC~1316 come from the recent infall of a small gas-rich galaxy. The 7.7~$\mu$m feature may be weakened by the X-ray radiation field in NGC~1316 that destroys the smaller PAHs. 
The 11.3~$\mu$m  PAH emission line from NGC~612 is as bright as that from NGC~1316, despite the-six-times greater distance. 

The steeper slope of NGC~1316 in the IRS spectrum may indicate a higher dust temperature than in NGC~612, likely associated with AGN heating. 
Both spectra exhibit high-excitation lines, such as [OIV] at 25.9$\mu$m, typically attributed to AGN. 

Could the synchrotron cores of the galaxies contribute significantly to the WISE fluxes and the \spitzer IRS spectra?   
Interestingly, the core of double-lobed radio galaxy Virgo~A at the center of M~87  
is detected across the entire electromagnetic spectrum, with a 24~$\mu$m MIPS flux of about 51~mJy  
\citep{2007ApJ...655..781S}. 
The non-thermal core and jet of M~87 were also detected in \wise images 
\citep{2013AJ....145....6J}. 
At 6~cm
(at $0\farcs5$ resolution), 
the core has a flux density of 2875~mJy 
\citep{2001ApJ...559L..87N}. 
For a similar spectral index, the flux density of Fornax~A and NGC~612 would be about 60 times weaker than at 6~cm, and therefore fainter than 1~mJy at 24~$\mu$m\footnote{
At 6~cm, the core of Fornax~A has a flux density of 26~mJy and that of NGC~612 38~mJy \citep{1993MNRAS.263.1023M}. 
}. This is much lower than what is measured in both galaxies' IRS spectra, which indicates that most of the emission in the long-wavelength part of the IRS spectra and in the \wise \wthree and \wfour images is not due to synchrotron emission from the nucleus. 
The jet of NGC~1316 was not detected in the \spitzer images either, and upper limits were set 
ranging between 0.39~mJy and 1.05~mJy between 3.6~$\mu$m and 24~$\mu$m   
\citep{2010ApJ...721.1702L}, 
in agreement with the expectations.   

The centers of both galaxies were observed at subarcsecond resolution in the MIR 
(the Sasmirala 
project\footnote{\tt http://dc.zah.uni-heidelberg.de/sasmirala}, 
\citealt{2014MNRAS.439.1648A}). 
The nucleus of NGC~612 was undetected, and an upper limit of 33~mJy at 12~$\mu$m was set, well below the measurement in the \spitzer IRS spectrum.
In NGC~1316, a flux density of $17\pm6$~mJy was measured at 12~$\mu$m, and a limit on the size of the emission of 38.1~pc was obtained (after correcting for our adopted distance). 
Those values show that in both galaxies the AGN is weak and the MIR emission is dominated by dust/star formation.  

\section{Resolved maps and component separation} \label{sec:composep}

The three main components that contribute to the mid-IR emission of NGC~1316 and NGC~612 (the stellar component, the AGN, and the dust) have different spatial and spectral shapes: 

-- The stellar component is extended and smooth and can be modelled using simple geometrical models; the spectrum is that of the Rayleigh-Jeans side of a black-body spectrum. 

-- The AGN, on the other hand, is point-like at the angular resolutions of \wise and can be modelled by a PSF; 
as a synchrotron source, its flux density varies with frequency as a power law, 
$S_\nu \propto \nu^\alpha$.  
The core of Fornax~A has a spectral index, $\alpha$, of $-0.52$ between 2.7 and 4.8~GHz and that of NGC~612 $-0.51$ 
\citep{1993MNRAS.263.1023M}. 
In Fornax~A, $\alpha$ decreases in the central jet to about $-1.7$ 
\citep{1984AJ.....89.1650G}. 

-- The dust in NGC~1316 appears very irregular, whereas in NGC~612 the dust lanes seen in extinction seem to be associated with a nearly edge-on warped disk. 
The mid-IR spectrum is a complex combination of spectral lines and thermal dust continuum \citep{2007ApJ...663..866D}; 
a good model of the thermal dust continuum is a superposition of modified black bodies at fixed temperatures (e.g. \citealt{2007ApJ...656..770S}). 

In this Section, we describe the two methods that we used to perform the component separation in the \wise images and present the results for each galaxy. 
In both methods, the \wone images serve as representatives of the starlight distribution.

\subsection{Methods}

\begin{table*}[t]
\centering
\caption{Best-fit models of the \wone (3.4~$\mu$m) image of NGC~1316.}
 \label{tab:fits-parameters}
\begin{tabular}{ccccccc}
\hline \hline
  & S\'ersic  & PSF 		&\vline &1st S\'ersic  & 2nd S\'ersic   & PSF \\
\hline
Image center (RA)  &  03:22:41.692  &  03:22:41.770 &\vline &  03:22:41.701  &  03:22:41.809  &  03:22:41.770\\
         (DEC)  &  --37:12:28.86  &  --37:12:27.54 &\vline &  --37:12:28.68  &  --37:12:32.09  &  --37:12:27.54 \\
Integrated magnitude  &4.87  &10.66 &\vline &    4.69  &    8.61  &   12.26\\
Flux density (mJy) &3456.9  &16.70 	&\vline &  4080.3  &   110.3  &     3.8\\
Effective radius, $R_{\rm eff}$ ($''$) &144.0&- &\vline &   248.0  &    37.7 &-\\
   S\'ersic index  &  6.15&- 		&\vline &    8.21  &    0.29 &-\\
Position angle, PA ($^\circ$)  &  52.7&- &\vline &    52.9  &    28.9 &-\\
 Axis ratio, $b/a$ &  0.70&- 		&\vline &    0.67  &    0.98 &-\\
Reduced chi-squared  &  21.8 & 		&\vline &    13.9\\
\hline
\end{tabular}
\end{table*}

The emission in both \wone (3.4~$\mu$m) and \wtwo (4.6~$\mu$m) is dominated 
by starlight. 
After examining the images, 
we decided to use the $W1$ images as representatives of the starlight in the galaxies. 
Indeed, although the \wone and \wtwo images are very close in wavelength, the noise is $\sim 36\%$ higher in $W2$; the galaxies are brighter in $W1$ by a factor of about 2, 
so the signal-to-noise is higher in the $W1$ image by a factor of about 2.5. 
The AGN is expected to be slightly fainter in $W1$ (by a factor of about 1.2 for NGC~1316, based on the estimates from the corresponding \spitzer images by Lanz et al. 2010) and is therefore less of a contaminant. 
Finally, we expect the PAH feature at 3.3~$\mu$m to be insignificant compared to the continuum. 

\subsubsection{Fitting method} 

In the first method we used GALFIT to model the starlight and the AGN. The residuals of the two-dimensional fits are a representation of the dust distribution in the galaxies.   

Our choice of starlight models was informed by the morphology of the galaxies, appreciable lower residuals and reduced chi-squared values.
For each galaxy, we tested two models of the starlight: 
for NGC~1316, a single S\'ersic and a double S\'{e}rsic model were used; 
for NGC~612, a single S\'{e}rsic model and the superposition of a S\'{e}rsic and an exponential disk model. As will be shown in the following sections, the double starlight models were better fits to the \wone images and were subsequently used to represent the starlight in the other bands. 

The AGN being very faint in $W1$, we had to use 
an iterative procedure with GALFIT to determine its location and flux, 
following the approach taken by 
\cite{2010ApJ...721.1702L}  
in their analysis of the \spitzer images of NGC~1316. 
We first determined the centers of the starlight in \wonep, then
used a starlight + PSF model to identify the location of the AGN in \wthreep.
We then kept these centers fixed and 
fitted a starlight + PSF model to the \wone image. 
The geometrical best-fit S\'ersic parameters from the \wone fit 
and the iteratively determined centers (of both the S\'ersic and the AGN) 
were held constant in the subsequent fits to the other images, while the 
amplitudes were left to float. 

\subsubsection{Scaling method}  

In the second method, we subtracted the starlight in the \wthree and \wfour images by simply scaling the \wone images by appropriate factors and subtracting them from the \wthree and \wfour images, after having smoothed them to the angular resolutions of \wthree and $W4$.  
This technique is commonly used to extract the non-stellar emission from  mid-IR images 
(e.g. \citealt{2004ApJS..154..253H},  
\citealt{2005ApJ...633..857D},  
\citealt{2007ApJ...663..866D}).  
The scaling factors were calculated from the spectral energy distribution of the stellar population of galaxies 
that are completely dominated by the evolved stellar population, 

\begin{eqnarray}
f_{\rm scaled~dust}(W3)= f(W3) - 0.158 f_{\rm smoothed~to~W3}(W1) 
\label{eqn:1} 
\end{eqnarray}
\begin{eqnarray}
f_{\rm scaled~dust}(W4)= f(W4) - 0.059 f_{\rm smoothed~to~W4}(W1) \, .
\label{eqn:2}
\end{eqnarray}

We used a 13~Gyr E-type galaxy from the GRASIL template \citep{1998ApJ...509..103S}. 

\subsection{NGC~1316} \label{sec:models} 

\begin{figure*}
  \includegraphics[width=6.cm]{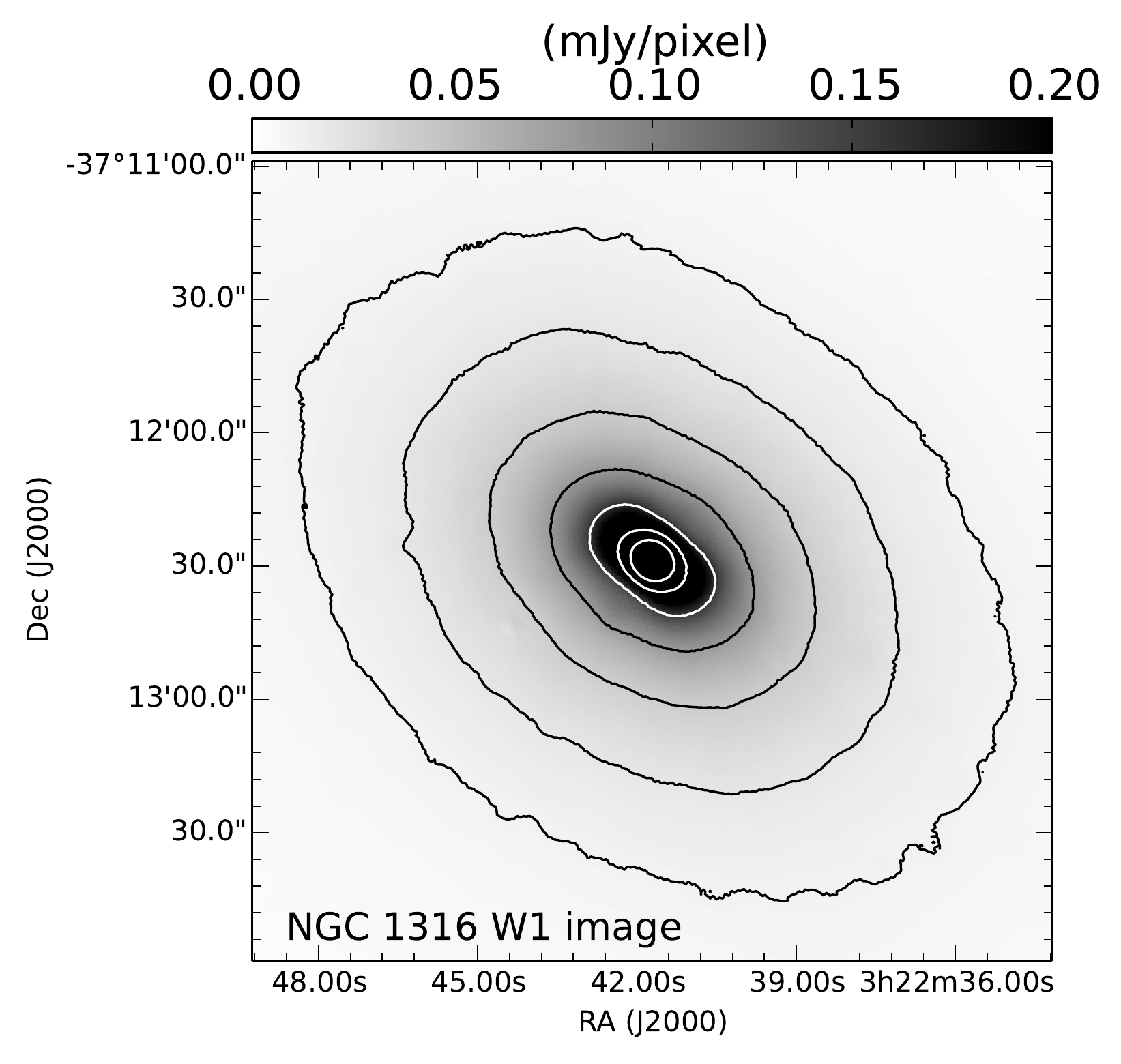}
  \includegraphics[width=6.cm]{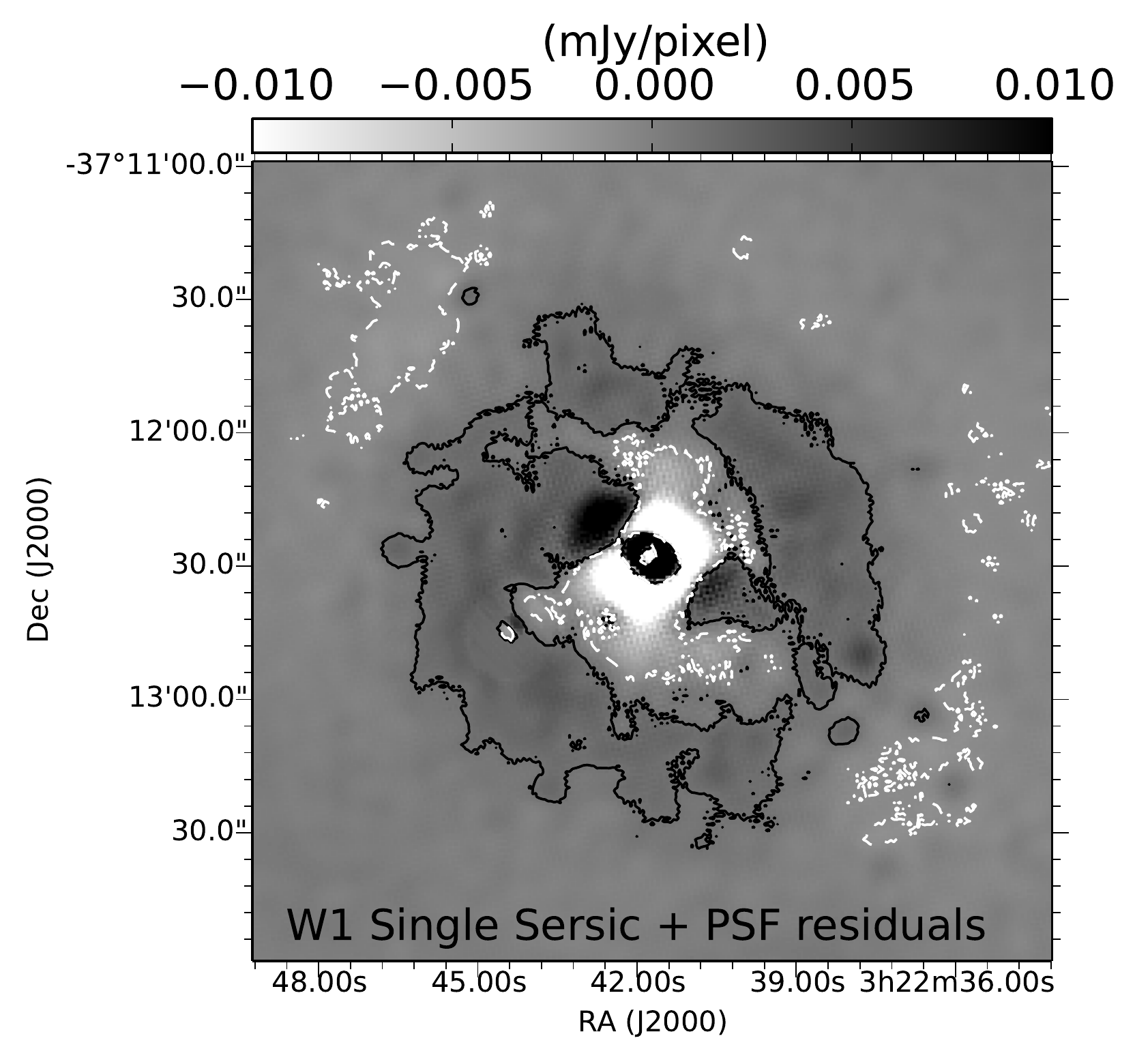}
  \includegraphics[width=6.cm]{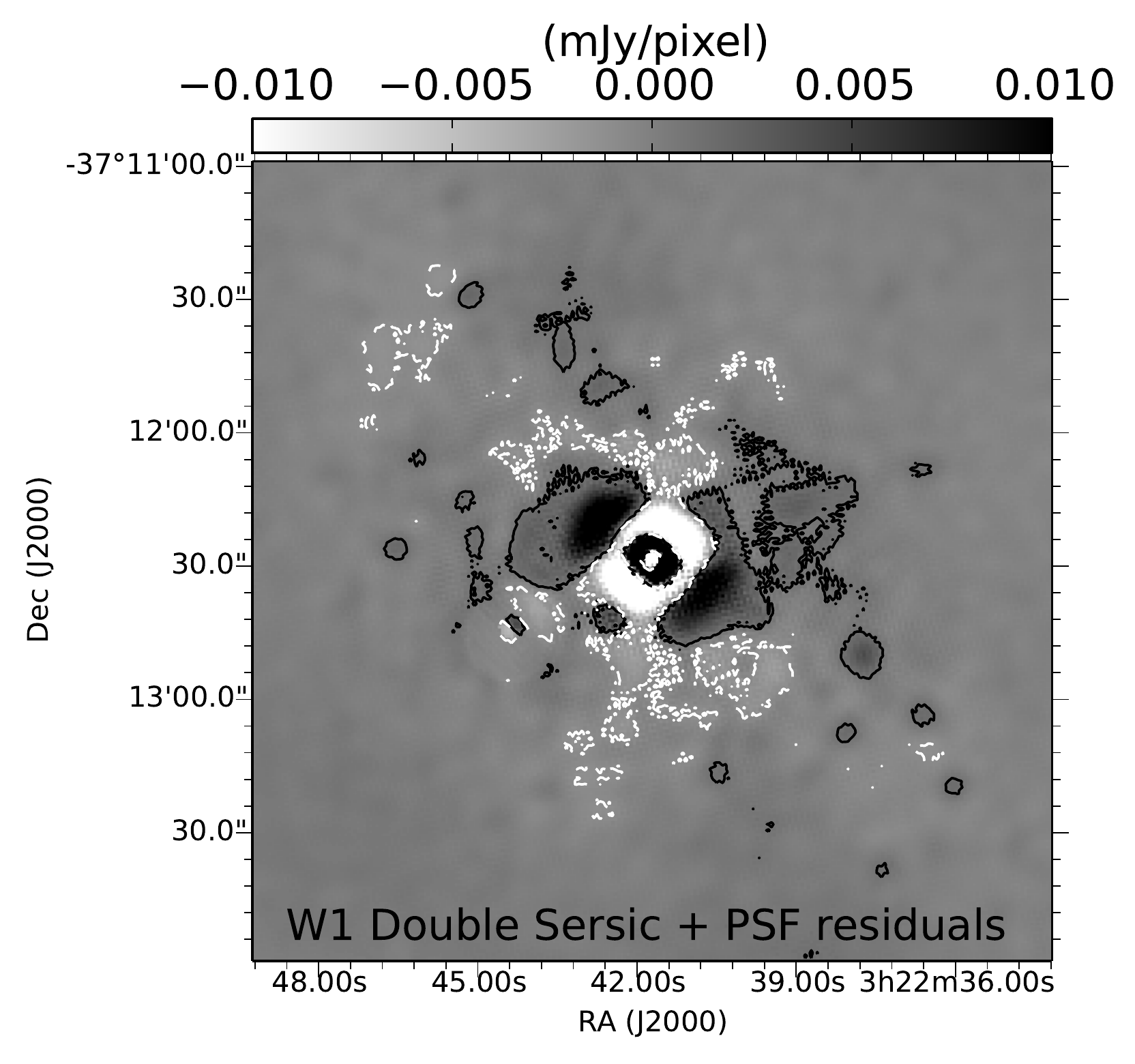}
\caption{
Characterizing the starlight distribution of NGC~1316: 
\wone image (left panel) and  
GALFIT residual maps for a single S\'{e}rsic + PSF model (middle panel), 
and a double S\'ersic model + PSF model (right panel). 
The contour levels of the \wone image increase from 10 to 640~$\mu$Jy/pixel in steps of a factor of 2; 
they are $\pm 1~\mu$Jy/pixel in the residual maps 
(positive contour in black, negative one in white). 
The best-fit parameters are in Table~\ref{tab:fits-parameters}. 
The double S\'ersic model gives lower residuals and will be used to
represent the starlight distribution of NGC~1316 in all WISE bands.
}
\label{figstarlightN1316}
\end{figure*}

\subsubsection{Starlight and AGN}
Figure~\ref{figstarlightN1316} 
shows the \wone image (left panel) and the residual maps 
for the two models considered, a single S\'ersic + PSF and a double S\'ersic + PSF (middle and right panels). 
The best-fit parameters are listed in Table~\ref{tab:fits-parameters}. 

The results can be directly compared to the ones obtained by 
\citet{2010ApJ...721.1702L} in their analysis of the \spitzer  
3.6~$\mu$m image of NGC~1316.   
They also used a single S\'ersic model + PSF;
the central coordinates and the geometrical parameters are in excellent agreement 
(they had a S\'ersic index of $6.07\pm0.10$, 
an effective radius of $146\farcs2\pm0\farcs6$, 
an axis ratio of $0.688\pm 0.004$ 
and a position angle of $54.0^\circ\pm0.2^\circ$). 
This is comforting given the similar wavelengths and comparable angular resolutions of 
the \spitzer 3.6~$\mu$m image and our enhanced-resolution \wone 3.4~$\mu$m image. 
The flux densities, however, are significantly higher in the \wise map, both for the S\'ersic model (about 3457~mJy for \wise versus 2390$\pm 70$ mJy for {\it Spitzer}) 
and for the central point source (16.7~mJy versus 4.67$\pm$0.22 mJy). 
Unfortunately \cite{2010ApJ...721.1702L} 
do not say whether their value for the S\'ersic model refers to the total flux or whether it was measured within a certain aperture.         
The WISE value in Table~\ref{tab:fits-parameters} refers to the total flux, out to infinite radius, while in Table~\ref{tab:ngc1316-sources} we give the flux within the 1-sigma isophotal radius of 761$''\simeq 12\farcm7$, 2.8~Jy, which is closer to the value quoted by \cite{2010ApJ...721.1702L}. 
Another explanation of the difference may be the background subtraction. 
Some of the diffuse emission may have been included in the background and subtracted.  
\cite{2010ApJ...721.1702L} 
mention that they used a $12'\times12'$ region for the analysis. 
The large field of view of WISE ($> 1$ degree) means that a clean background measurement is possible. 

The residual map for the single S\'ersic + PSF model shows a positive ring-like feature 
at a radius of about 30$''$, with a width of about $20''$. 
In the more central region, on the other hand, the model slightly oversubtracts the emission and then undersubtracts it (black and white regions). 
Very similar features can be seen in the residual \spitzer 3.6~$\mu$m map of 
\cite{2010ApJ...721.1702L}. 
Using a double S\'ersic model + PSF removes completely the ring-like structure, whereas the very central region remains the same. 

Compared to the single S\'ersic + PSF model, the number of parameters has increased from 10 to 17, 
for a total number of data points of 525$^2$. 
The reduced chi-square values have decreased significantly (see Table~\ref{tab:fits-parameters}), 
which indicates that the double S\'ersic model is a better representation of the stellar light in NGC~~1316. 
It will therefore be used in the rest of the analysis.  

\subsubsection{Non-stellar emission}

\begin{figure*}[hbt!]
\includegraphics[width=8.8cm]{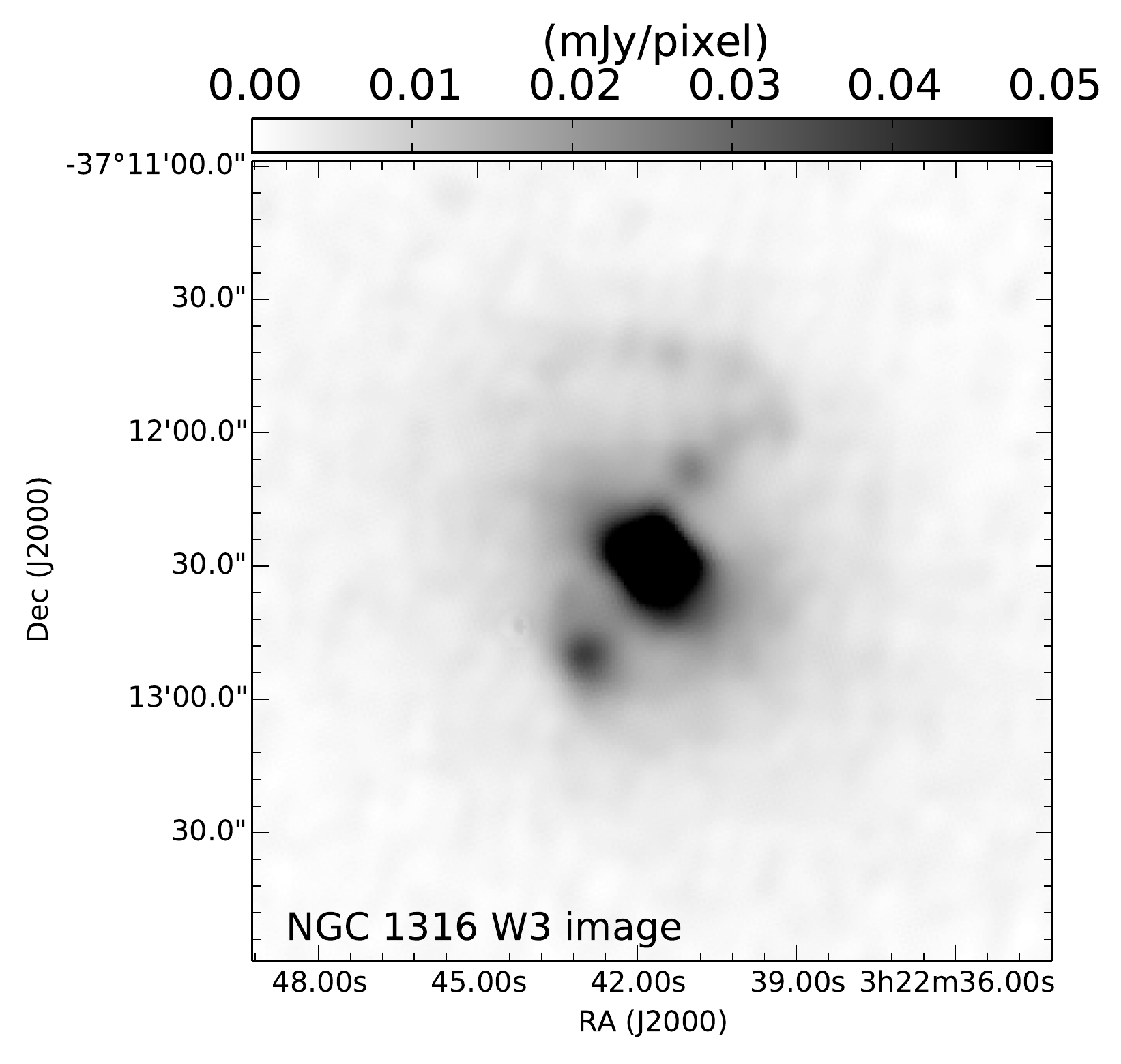} 
\includegraphics[width=8.8cm]{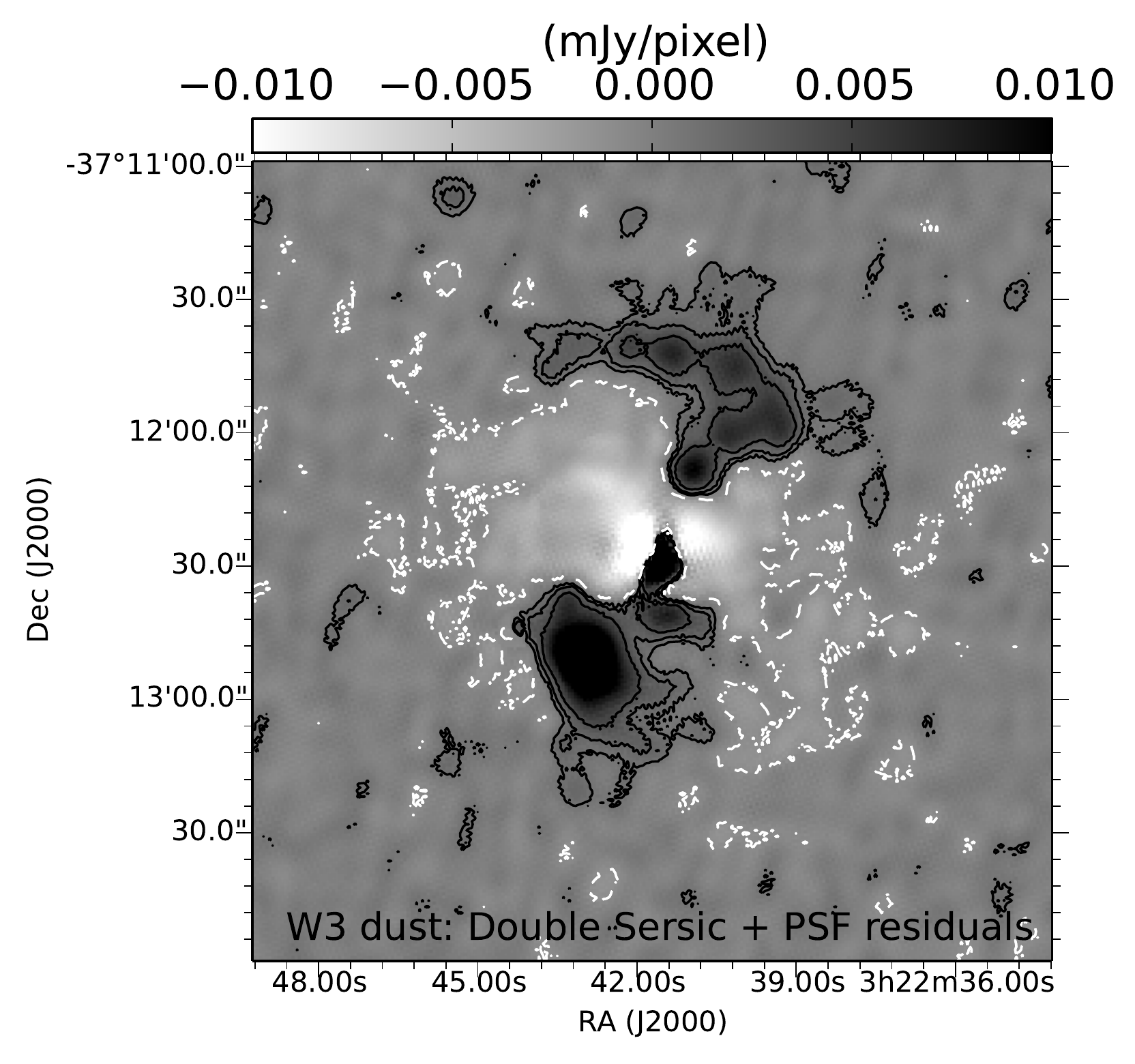}
\caption{
{\it Left panel:} \wthree (12~$\mu$m) image. 
{\it Right panel:} Residuals from the double S\'ersic + central PSF model fitting.  
The contour levels are $- 1~\mu$Jy/pixel (in white) and $+1$, $+2$ and $+4~\mu$Jy/pixel (in black). 
} 
\label{figN1316W3galfit}
\end{figure*}

\begin{figure*}[hbt!]
\includegraphics[width=8.8cm]{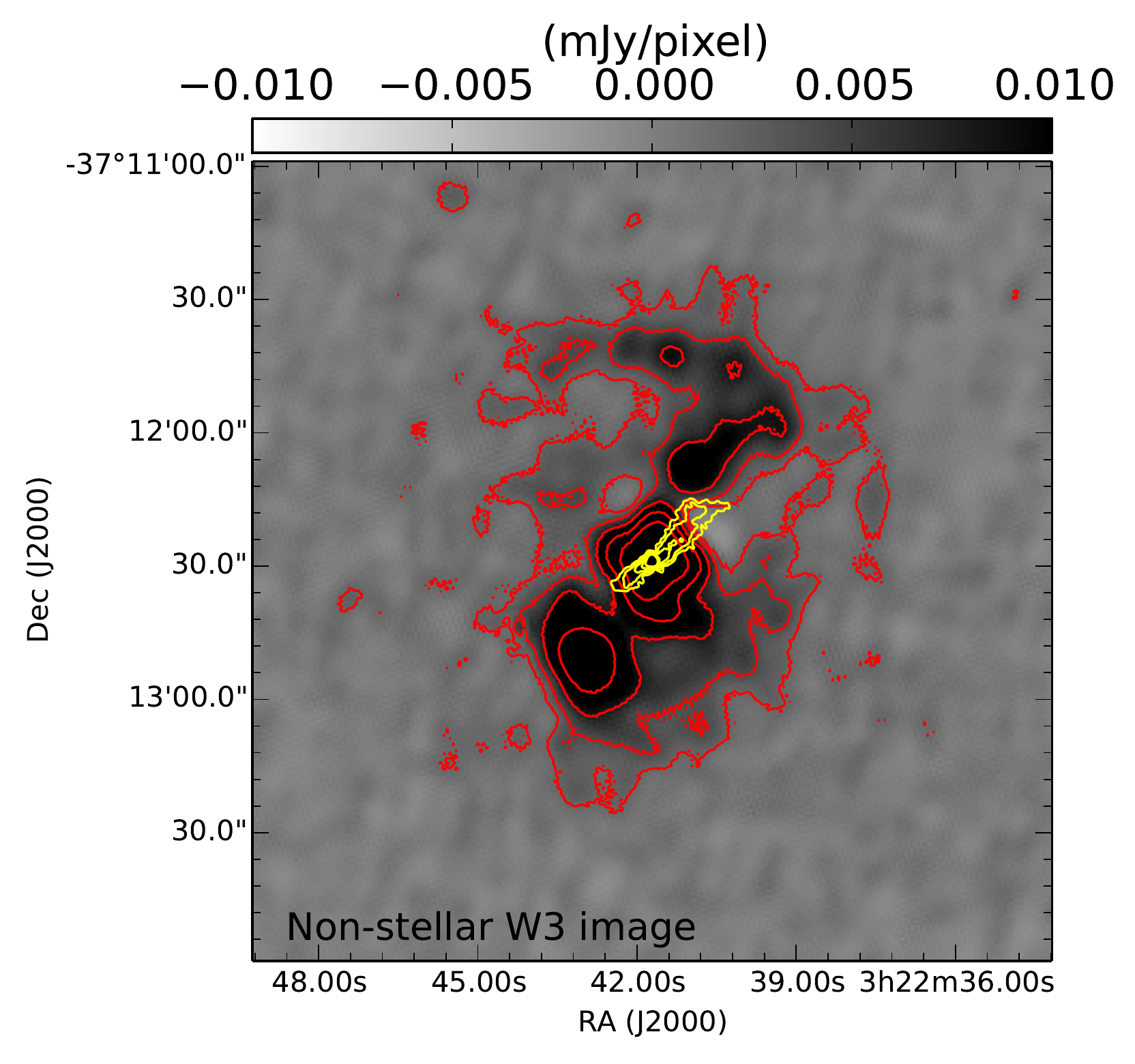} 
\includegraphics[width=8.8cm]{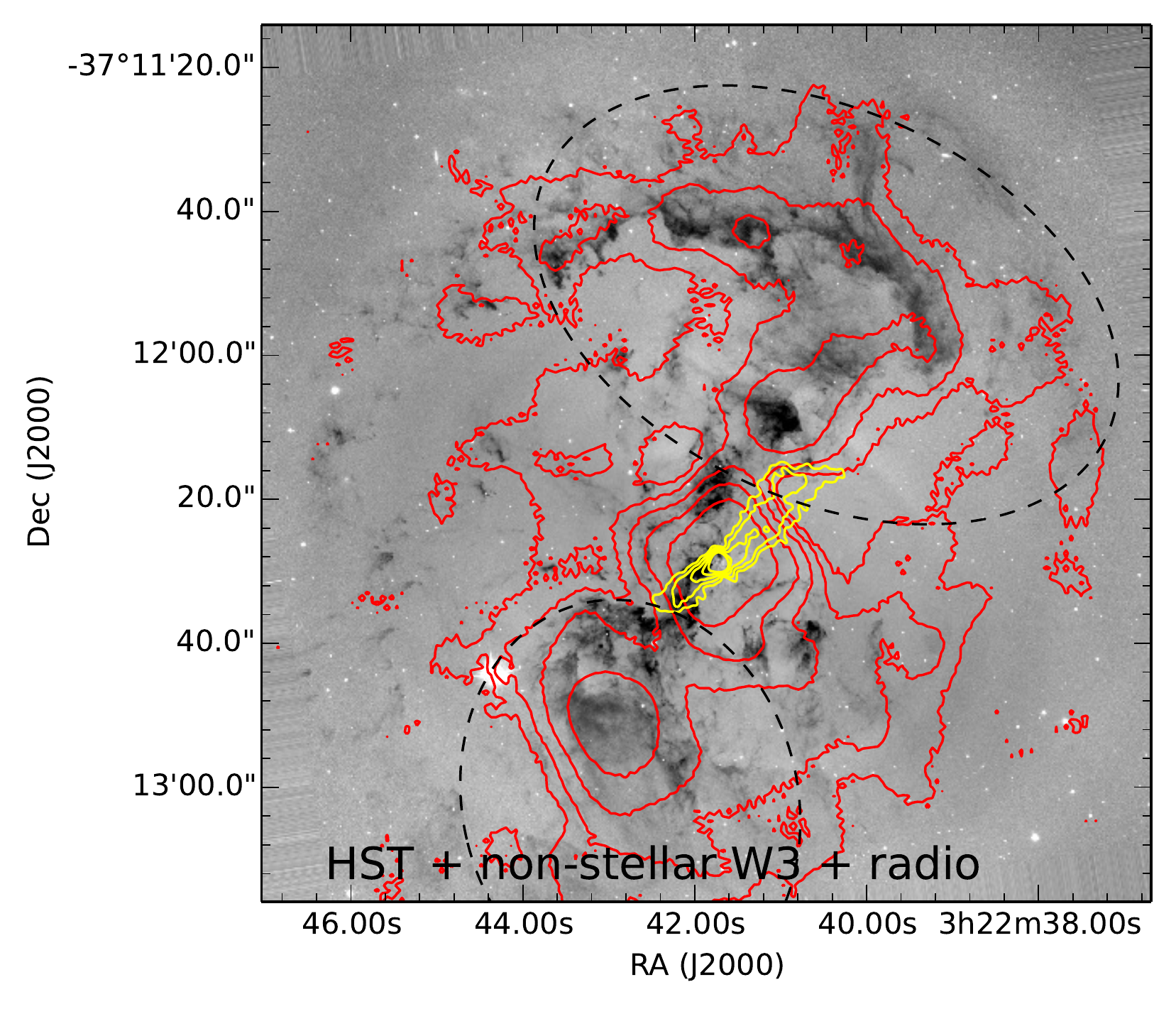} 
\caption{ 
In both panels, the red contours show the \wthree (12~$\mu$m) 
non-stellar image of NGC~1316 obtained by the scaling method and the yellow contours the inner radio jet (as in Fig.~\ref{figHSTDSS}). 
The red contours increase from 2 to 32~$\mu$Jy/pixel by a factor of two. 
The substructures agree with those seen in the GALFIT subtraction (Fig.~\ref{figN1316W3galfit}) 
within 1~$\mu$Jy/pixel, except in the central regions which were oversubtracted by GALFIT. 
{\it Left panel:} Non-stellar \wthree image in grey-scale. 
{\it Right panel:} HST image in grey-scale. 
The dashed ellipses show two regions to the NW and SE in which our WISE dust measurements will be compared to the \spitzer measurements \citep{2010ApJ...721.1702L}. 
} 
\label{figN1316W3scaled}
\end{figure*}

Figure~\ref{figN1316W3galfit} (left panel) shows the \wthree image of NGC~1316. 
The residual image obtained from the fitting method to subtract the stellar component and a possible AGN is shown in the right panel. 
This image reveals clearly the dusty features to the NW (the mushroom-like structure with an azimuthal arc) and SE of the nucleus seen in the HST image and  
known from earlier mid-IR studies 
(e.g. 
\citealt{2004A&A...416...41X}, 
\citealt{2005ApJ...622..235T}, 
\citealt{2010ApJ...721.1702L}, 
\citealt{2014arXiv1409.2474D}). 

The limitations of the fitting method are visible in the central region, where emission has clearly been oversubtracted (there are negative values).
The S\'ersic models do not perfectly represent the stellar light, as indicated earlier in the fits to W1. 
The PSF component also clearly overestimated the central flux. 
The central region of NGC~1316 contains dust patches very close to the AGN and 
those components are mixed at the 
resolution and sensitivity of the \wise images.

Fig.~\ref{figN1316W3scaled} shows the image of NGC~1316 obtained using the scaling method to remove the starlight contribution (traced by $W1$) to the \wthree image. 
The scaling method yields more extended emission than the fitting method. 
It is possible that the S\'ersic models included part of the diffuse dust emission, since their amplitudes were left to vary freely during the fit (while the geometrical parameters were fixed). If the dust distribution partly follows that of the starlight, then 
the two components cannot be separated by the fitting method and part of the dust emission may be subtracted. 

In order to check the sensitivity of the features to the adopted scaling factor of 0.158 for the \wone image, we slightly varied this factor. Increasing it only slightly (to 0.17) produces a negative ring around the center, which indicates that this factor, which was estimated from SED modeling of dust-free elliptical galaxies, must be correct within 10\% for NGC~1316. Note that 
the scaling method does not subtract the AGN in \wthree  
since it is much weaker in \wone than in \wthreep. 

In order to quantify the dust emission in NGC~1316 and compare to other estimates, we measured the flux density in the two 
elliptical regions defined by \cite{2010ApJ...721.1702L} 
and shown in Fig.~\ref{figN1316W3scaled} (right panel). 
Interestingly, the two regions have very comparable flux densities in \wthreep: 
$25.3 \pm 5.0$~mJy in the NW region and 
$26.0 \pm 5.1$~mJy in the SE one. 
At 5.8~$\mu$m and 8.0~$\mu$m, 
\cite{2010ApJ...721.1702L} had measured 
$\sim 6.5$~mJy and 13.8~mJy in the NW region and  
$\sim 4.2$~mJy and 18.8~mJy in the SE region.  
Without a spectrum it is not possible to be certain that the higher 12~$\mu$m to 8~$\mu$m flux ratio in the NW region is due to a different 11.3~$\mu$m to 8~$\mu$m PAH flux ratio. The NW region is known to be about 2.5 times richer in molecular gas \citep{2001A&A...376..837H}. 
This is also where the inner jet ends abruptly and it may have affected the physical conditions of the interstellar medium. 

The methods could not be applied successfully to the \wfour image of NGC~1316 because of the very bright nucleus that causes artifacts in the deconvolution in the form of negative regions and sidelobes. The global properties, however, were characterized and a total flux of about 342~mJy was estimated 
(Table~\ref{tab:ngc1316-sources}). This is slightly lower that what \cite{2010ApJ...721.1702L} 
had measured in their \spitzer 24~$\mu$m image ($400\pm 12$~mJy for the S\'ersic component 
and $60.8\pm1.8$~mJy for the nuclear source). Since the \wfour image is of limited interest compared to the \spitzer 24~$\mu$m image, it is not further discussed.

\subsection{NGC~612} 
\subsubsection{Starlight}

\begin{table*}[t]
\centering
\caption{Best-fit models of the \wone (3.4~$\mu$m) image of NGC~612.}
 \label{tableN612W1}
\begin{tabular}{cccccc}
\hline \hline
  			&S\'ersic       &\vline	&S\'ersic  &+ &Exponential disk\\
\hline
Image center (RA)  	&01:33:57.634	&\vline	&01:33:57.618	& &01:33:57.637\\
             (DEC)  	&--36:29:35.36	&\vline &--36:29:35.21	& &--36:29:35.26\\ 
Integrated magnitude  	&9.34		&\vline	&9.84		& &10.10\\    
Flux density (mJy) 	&56.3		&\vline	&35.5		& &28.0\\
Effective radius, $R_{\rm eff}$ ($''$)&9.8 &\vline &22.1	& &-\\
S\'ersic index  	&2.75  		&\vline	&1.68		& &-\\
Disk scale length $R_{\rm s}$ ($''$)&-	&\vline	&-		& &3.2\\
Position angle, PA ($^\circ$)  	&--4.2	&\vline	&--15.1		& &--1.0\\
Axis ratio, $b/a$	&0.49		&\vline	&0.73		& &0.34\\
Reduced chi-squared  	&13.7		&\vline	&		&4.6\\
\hline
\end{tabular}
\end{table*}

\begin{figure*}
\includegraphics[width=6.25cm]{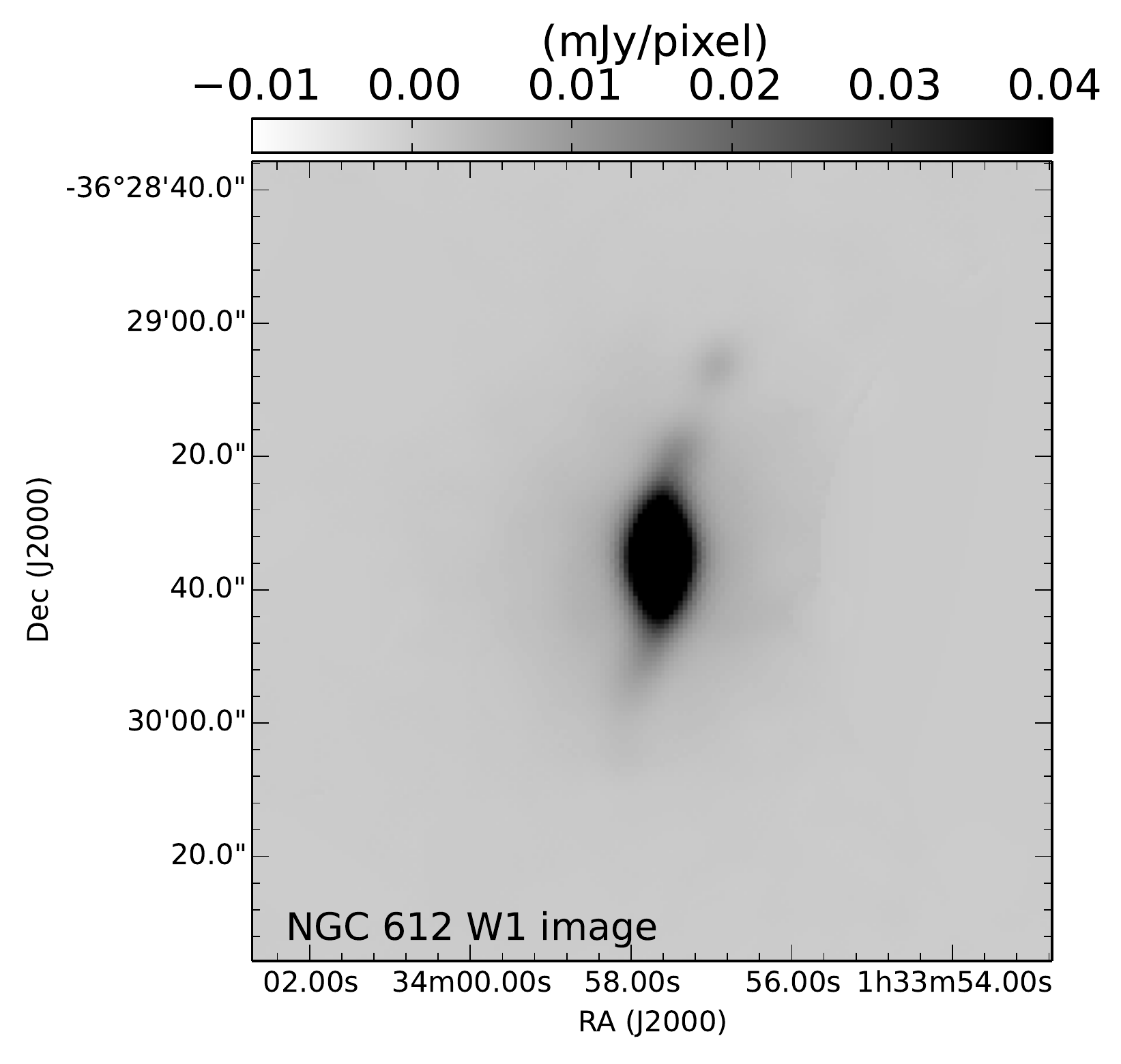}
\includegraphics[width=6.25cm]{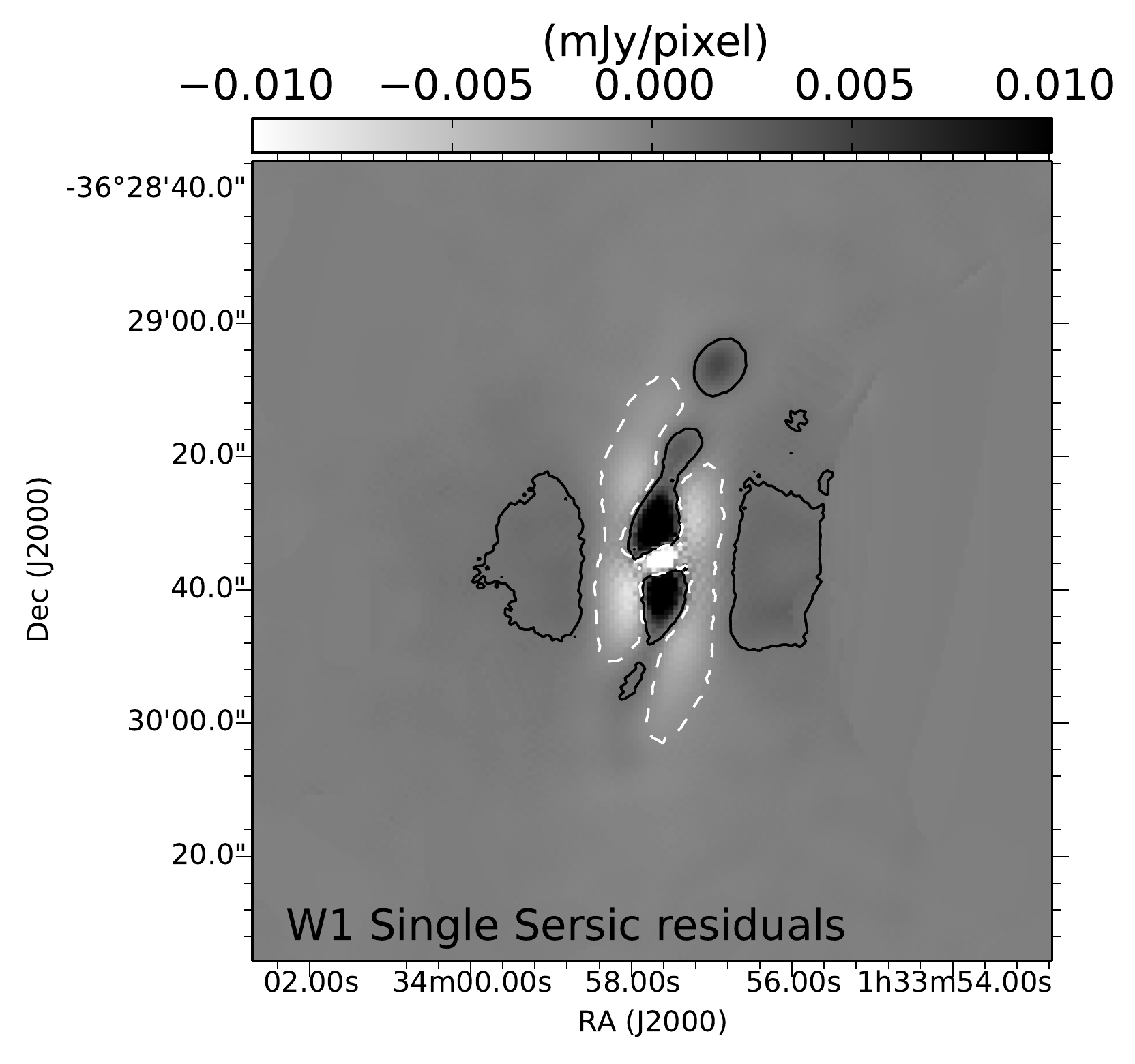}
\includegraphics[width=6.25cm]{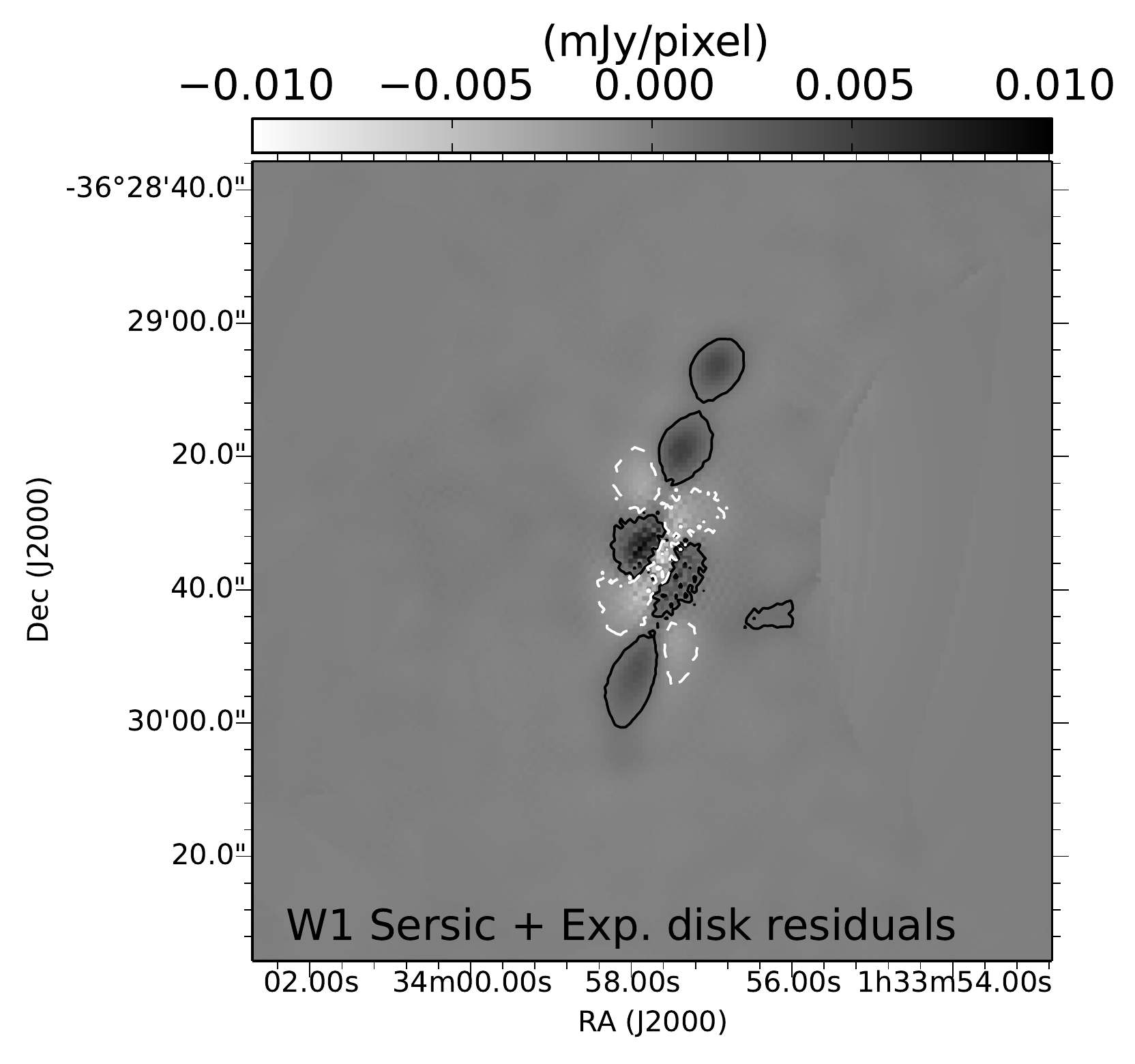}
\caption{
Characterizing the starlight distribution of NGC~612: \wone image (left panel) and GALFIT residuals  
using a single S\'{e}rsic model (middle panel) 
and a S\'ersic + exponential disk model (right panel). 
The contour levels are $\pm 1\mu$Jy/pixel 
(positive contour in black and negative one in white) 
and the best-fit parameters are given in Table~\ref{tableN612W1}.
The S\'ersic + exponential disk model gives lower residuals and will be used to represent the starlight distribution of NGC~612 in all WISE bands.
}
\label{figstarlightN612}
\end{figure*}

For NGC~612 we used a single S\'ersic model and a S\'ersic + exponential disk to represent the starlight. 
We also included a PSF to model a possible central source, but the fitting procedure did not identify such a component in any of the bands.  
The $W1$ image of NGC~612 and the residuals of the fits are shown in Figure~\ref{figstarlightN612}. The best-fit parameters are given in 
Table~\ref{tableN612W1}.  

Adding an exponential disk to the S\'ersic model increases the number of parameters from 10 to 16, 
for the same number of data points of 350$^2$. The residual map and the chi-square are better for this two-component model. 
Note, however, that the warp in NGC~612 cannot be captured by those simple models. 

The effective radius of NGC~612 in the single S\'ersic model of the \wone image is slightly lower than what was found in the $R$-band 
($9\farcs8$ (6~kpc) instead of $16\farcs7$;
\citealt{1996MNRAS.282...40F}, \citealt{2000A&A...353..507G}) 
and in the $i$-band ($17\farcs7$; \citealt{2001A&A...375..791V}). 
Our values can be compared more directly to those found 
in the $K$-band by \cite{2010MNRAS.407.1739I} 
who also included an edge-on disk and 
obtained an $R_{\rm eff}$ of $20\farcs25$ for a fixed S\'ersic index of 4 
and a disk scale-length comparable to ours, $3\farcs4$ (2~kpc).

\subsubsection{Non-stellar emission}

\begin{figure*}
\includegraphics[width=6.25cm]{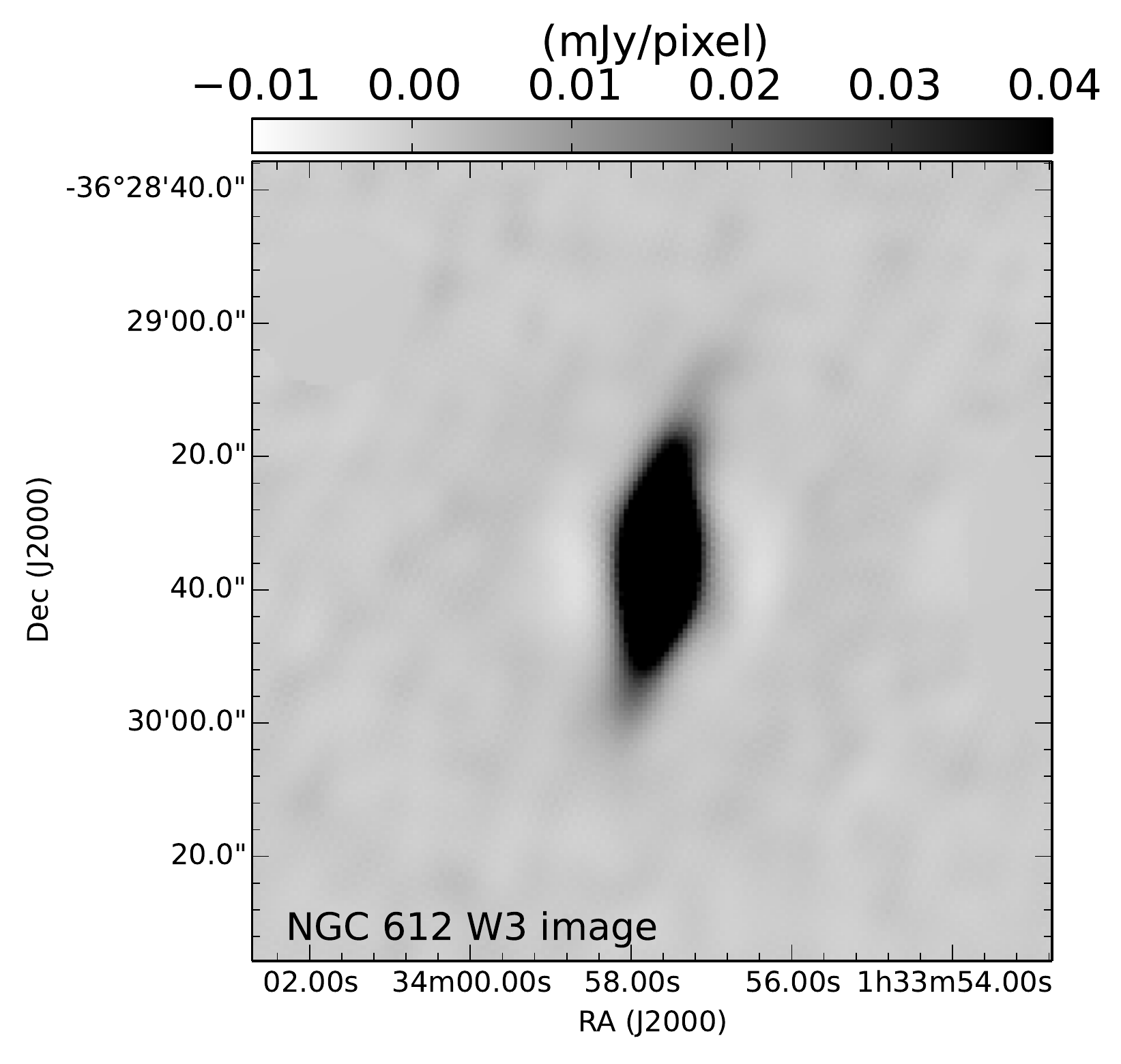}
\includegraphics[width=6.25cm]{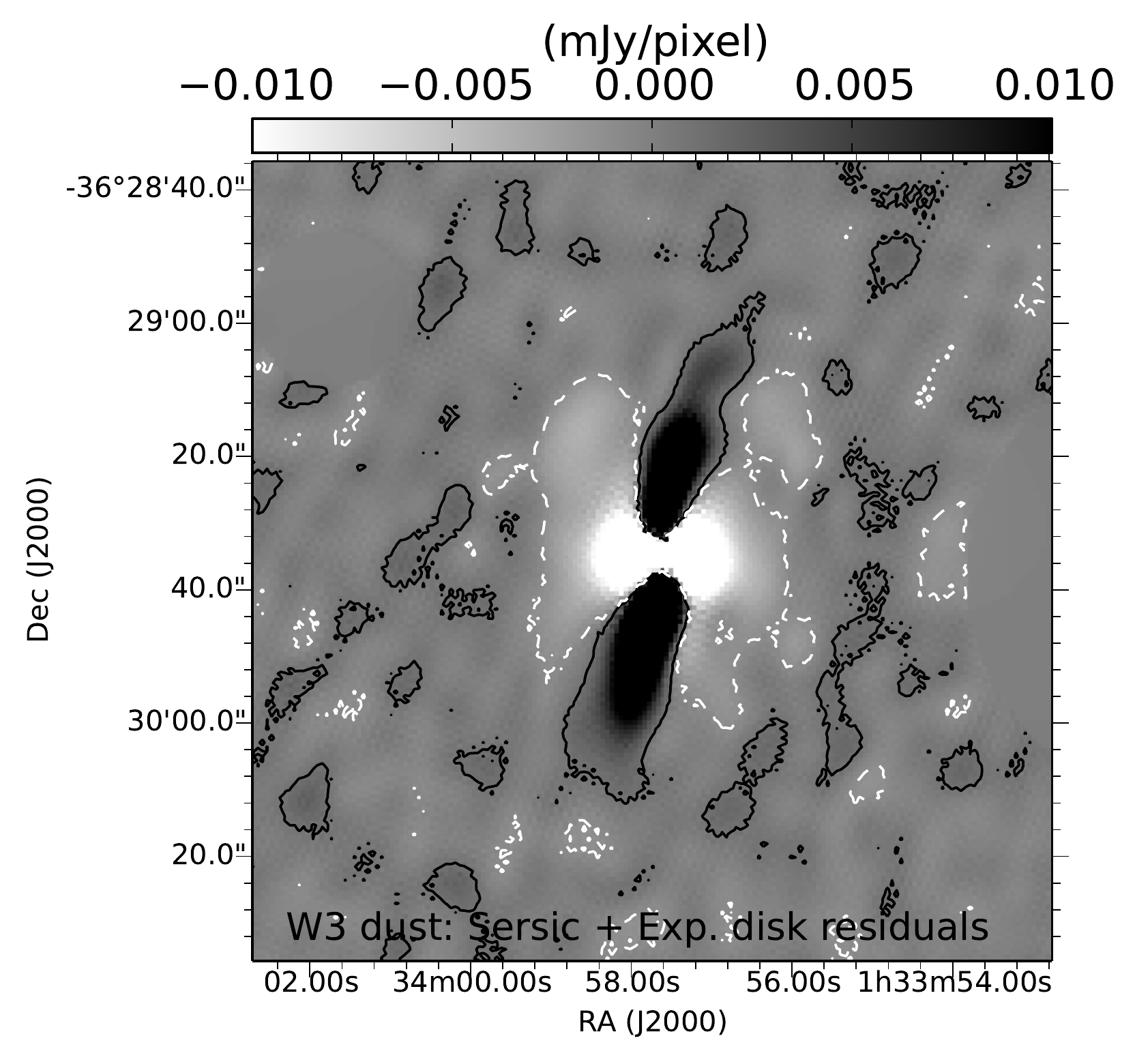}
\includegraphics[width=6.25cm]{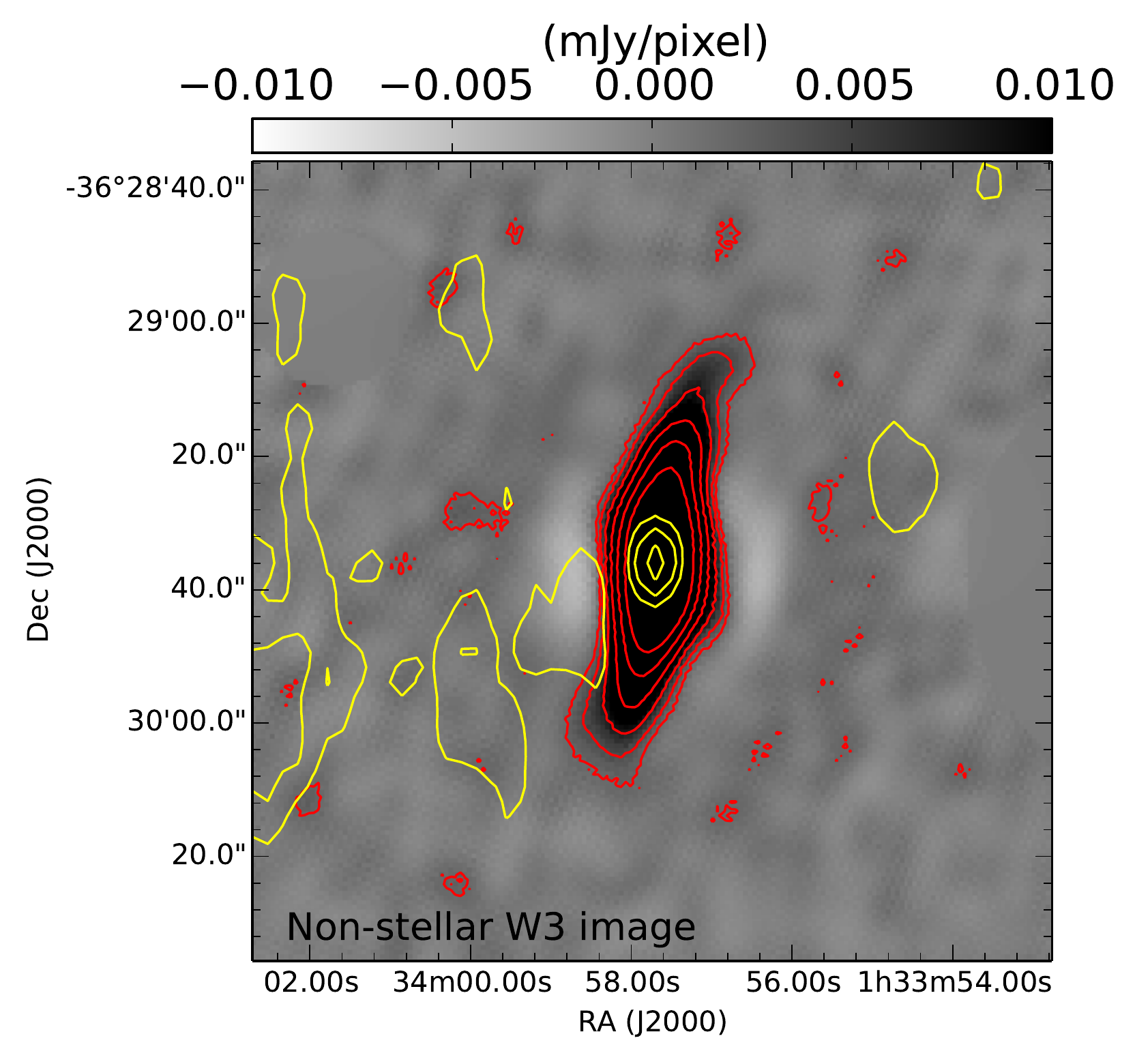} \\ 
\includegraphics[width=6.25cm]{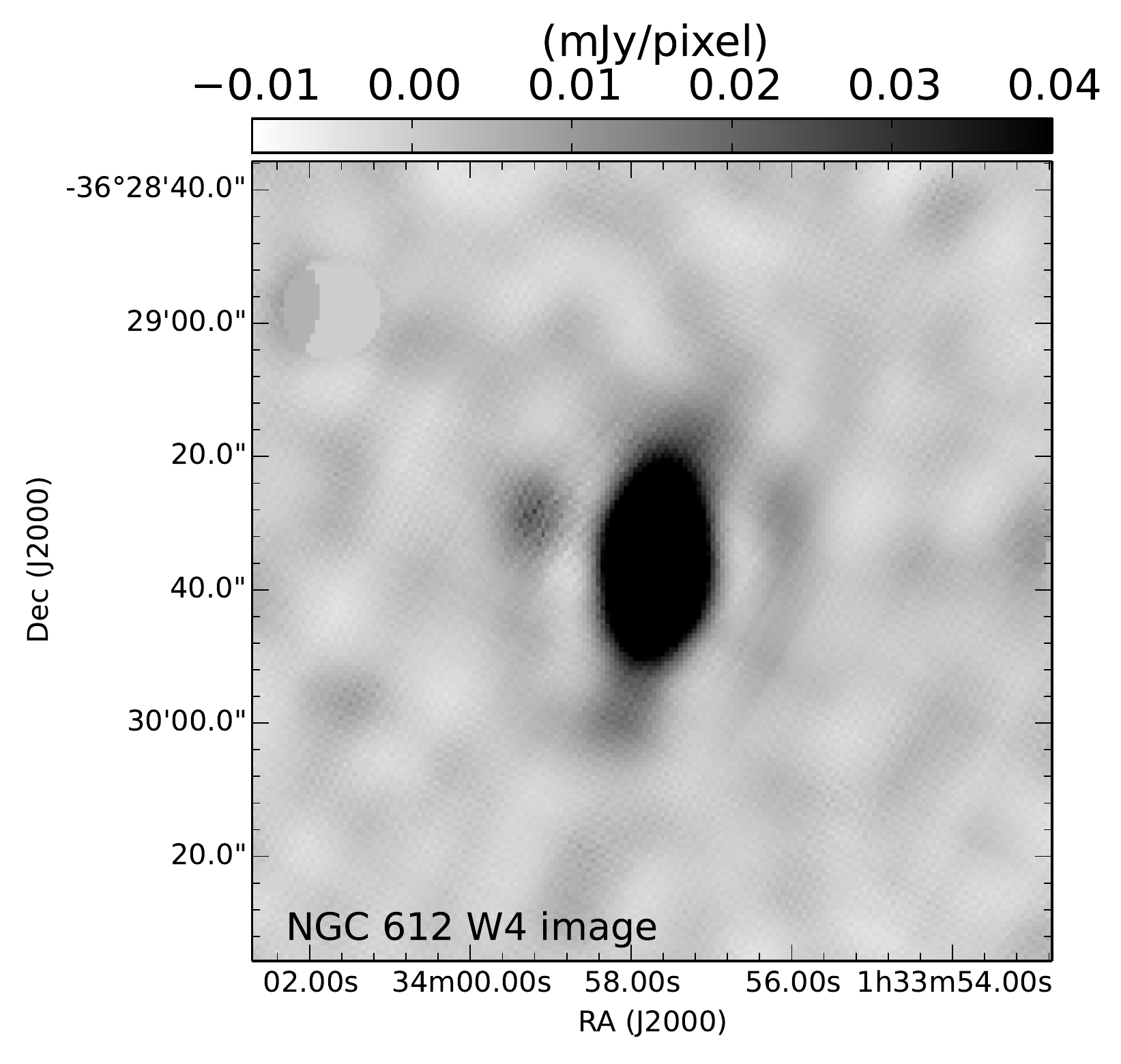}
\includegraphics[width=6.25cm]{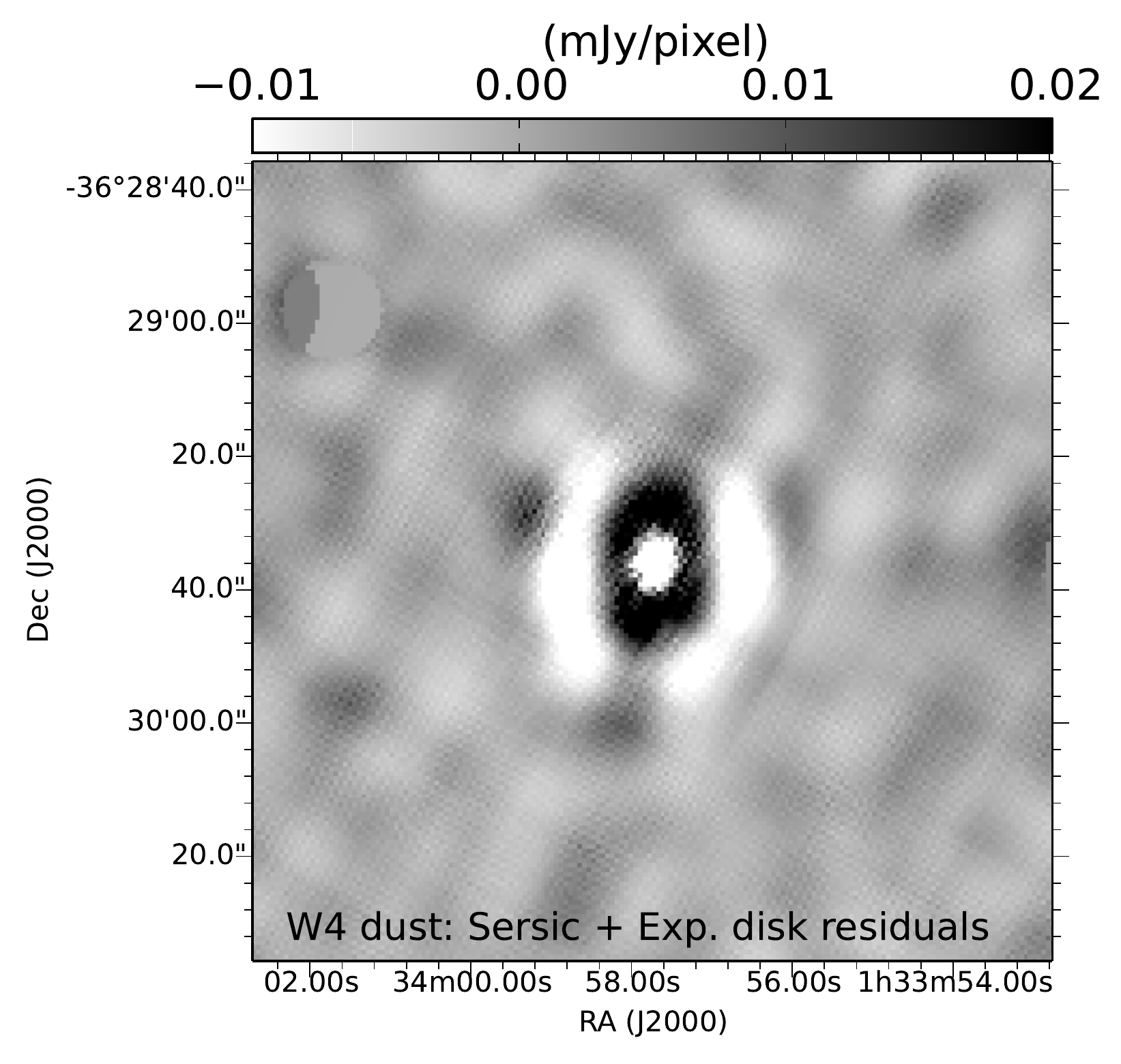}
\includegraphics[width=6.25cm]{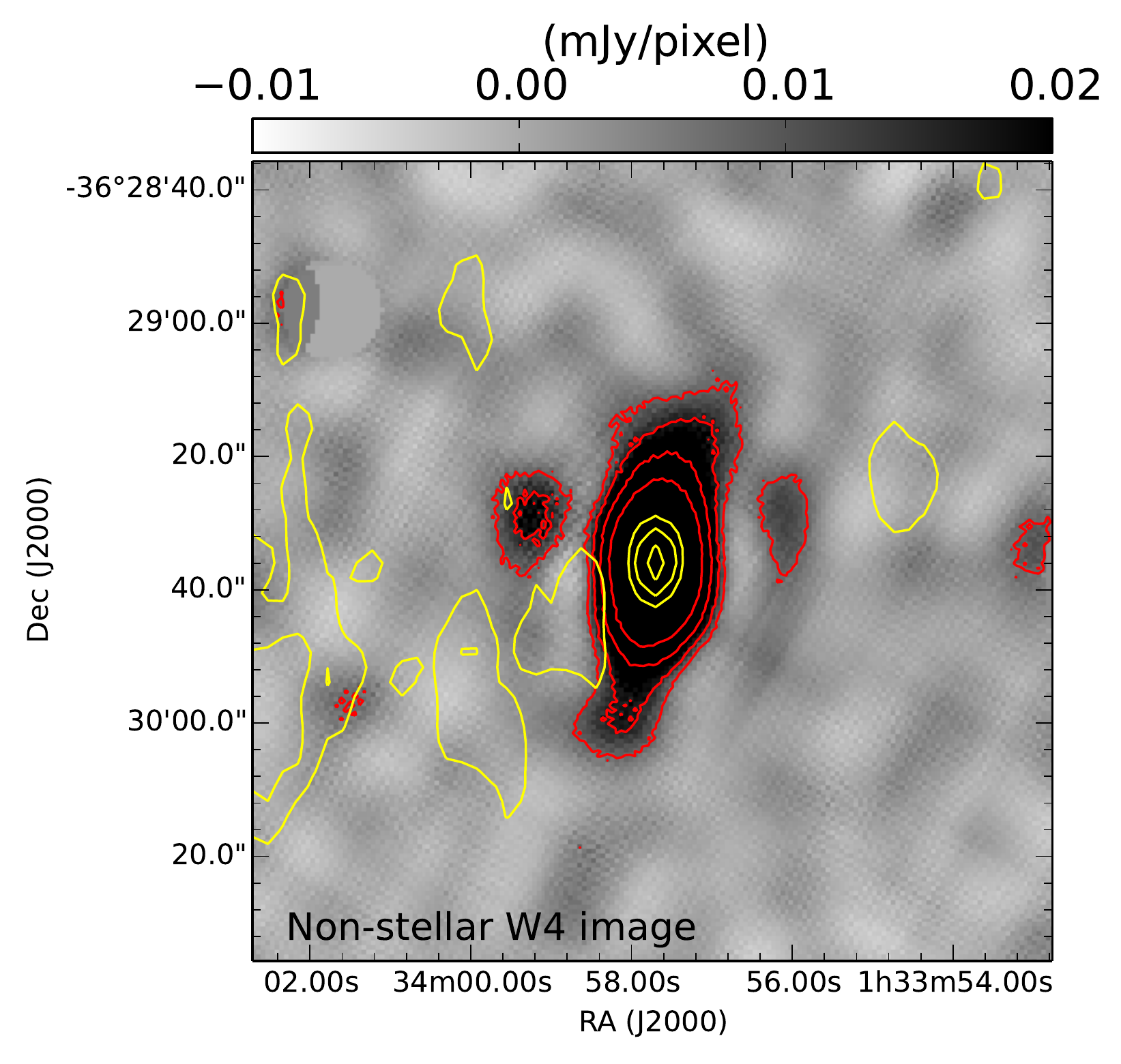}
\caption{
Results of the analysis of the \wthree (12~$\mu$m) and \wfour (22~$\mu$m) images of NGC~612. 
The first row shows the \wthree image (left panel), 
the residuals from 
the S\'ersic + exponential disk model fitting (middle panel), 
and the image obtained from the scaling method (right panel). 
The contours in the image of the \wthree residuals of NGC~612 (middle row) are $\pm 1 \mu$Jy/pixel 
(positive contour in black, negative one in white). 
The second row shows the corresponding images for \wfourp. 
The red contour levels increase by a factor of 2 from 2 to 64~$\mu$Jy/pixel for the \wthree image and  
 from 8 to 64~$\mu$Jy/pixel for the \wfour image. 
The yellow contours show the radio emission and are the same as in Fig.~\ref{figHSTDSS}. 
} 
\label{figdustN612}
\end{figure*}

\begin{figure}
\includegraphics[width=8.8cm]{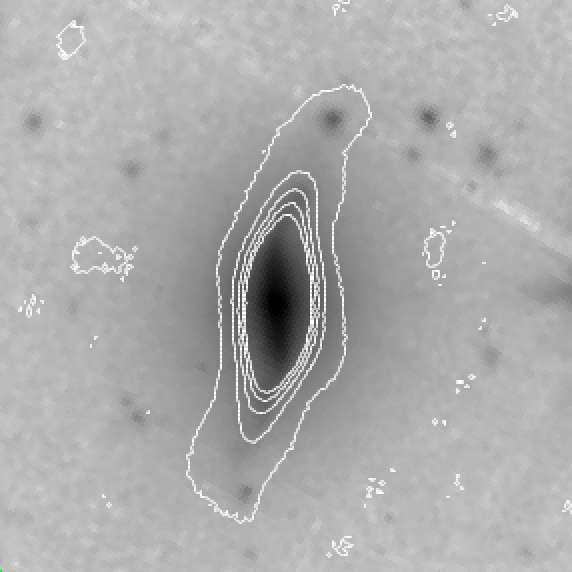}
\caption{
Archival \spitzer 4.5~$\mu$m image of NGC~612 and superimposed contours of the \wthree 12~$\mu$m scaled image (same contours as in Fig.~\ref{figdustN612}).
The warped disk is visible.  The size of the image is $1.5'\times1.5'$.
}
\label{figN612spitzerW3}
\end{figure}

Figure~\ref{figdustN612}  
shows the \wthree and \wfour images (left panels) and the maps of the non-stellar emission obtained using both methods 
(fitting method in the middle panel and scaling method in the right panel). 
The \wise images of NGC~612 clearly reveal a warp, best seen in the scaled \wthree image.  
The structure of the dust lane is difficult to see in optical images such as the DSS image shown in Fig.~\ref{figHSTDSS}. The warp, however, appears also in archival \spitzer images\footnote{ 
Project 10098, `Warm Spitzer Imaging of NuSTAR fields', PI Daniel Stern}, as shown in the overlay of the scaled \wthree image on the \spitzer 4.5~$\mu$m displayed in Fig.~\ref{figN612spitzerW3}. 
It seems to extend to much larger radii, especially to the North, in the HI image of Emonts et al. (\citeyear{2008MNRAS.387..197E}; their Fig.~7). 

The two methods give different results: the fitting method removes much more flux than the scaling method: the fit to \wthree yielded a flux of 12.2~mJy for the S\'ersic component, and of 99.8~mJy in the exponential disk; if we scale the components fitted in \wone using our adopted scaling factor, then the fluxes from the stellar population in \wthree are only 5.6~mJy in the S\'ersic and 4.4~mJy in the exponential disk. 
This is understandable because the dust in this galaxy, unlike in NGC~1316, follows the bulk of the stars. The dusty component is therefore largely included in the fit (especially in the exponential disk) and gets subtracted. The relative merits and limitations of both methods will be discussed in Sect.~\ref{SectCompa}.

\section{Colors, stellar masses and star formation rates} \label{sec:SMASS-SFR}

\subsection{WISE colors} \label{sec:band-colors}

\begin{table*}[t]
\centering
\caption{Colors in the WISE bands.} 
\begin{tabular}[t]{l c c c c }
\hline \hline
Galaxy & $[W1-W2]$ & $[W2-W3]$ & $[W1-W3]$& $[W3-W4]$\\ 
		& (mag)& (mag) &(mag) & (mag)\\
\hline
NGC~1316 & $-0.022\pm0.016$ & $0.623\pm0.016$ &$0.601\pm0.016$ & $1.261\pm0.020$\\ 
NGC~612 & $0.132\pm0.016$ & $3.069\pm0.017$ &$3.200\pm0.016$ & $1.621\pm0.020$\\
\hline
\end{tabular}
\tablefoot{The colors were estimated in matched apertures (the \wtwo isophotal apertures). 
}
\label{tab:wise-colors}
\end{table*}

It is interesting to examine how the WISE colors of our two galaxies (presented in Table~\ref{tab:wise-colors}) 
compare to those of larger galaxy populations. 
The WISE \wonep-\wtwo vs. \wtwop-\wthree color diagram has been investigated by a number of authors  
to study statistically large galaxy populations and their evolutionary states
(e.g. \citealt{2013AJ....145....6J}, \citealt{2014ApJ...782...90C}). 
The color-color diagram crudely separates morphological types, notably disk (star-forming) from elliptical (passive, or quenched) and from AGN-dominated. 
\cite{2013ApJ...772...26A} 
found that large \wonep-\wtwo colors (larger than 0.8~mag) can be used to identify bright AGN. 
\cite{2014ApJ...794L..13A} 
identified a clear transition zone in $u-r$ versus \wtwop-\wthree diagrams between the loci of early-type galaxies and late-type galaxies. 
\cite{2014MNRAS.438..796S} 
saw a strong dichotomy in the WISE colors of FR~I and FR~II radio sources, the FR~I being found in WISE early-type galaxies and 
the FR~II in late-type galaxies 
(FR~I radio galaxies have diffuse radio lobes with an outwards declining brightness and are typically fainter than 
FR~II sources that have brighter edges with hot spots; 
\citealt{1974MNRAS.167P..31F}). 

The two galaxies are found on different sides of the infrared transition zone. 
NGC~1316 has mid-infrared colors that are dominated by the old stellar population host galaxy, with only minimal signature of an AGN (and hence the colors are consistent with an evolved luminous population), while NGC~612 has very red W2-W3 colors, consistent with active star formation (as well as an AGN, noted by the warm \wonep-\wtwo color) while also maintaining a bulge component (most pronounced in the near-infrared, exhibiting an S0 morphology).

In the WISE color-color diagram of local radio galaxies (Fig.~13 of \citealt{2014MNRAS.438..796S}), 
NGC~1316 appears in the group of FR~I radio galaxies.  
NGC~1316 does not have significant radio hot spots and the radio lobes are not edge-brightened, so it can be classified as an FR~I radio galaxy. 
NCC~612 is clearly outside the region dominated by FR~I galaxies with its much redder \wtwop-\wthree color, 
consistent with active star formation; 
with a \wonep-\wtwo color of about 0.13~mag, NGC~612 appears at the lower end of the \wonep-\wtwo distribution of FR~II radio galaxies (note that the FR~II galaxies spread over a large range of \wonep-\wtwo colors, between about 0 and 1.2 mag): NGC~612 is not as red in its \wonep-\wtwo color as most FR~II galaxies, which are associated with spirals and quasi-stellar objects/Seyfert galaxies. 
Interestingly, NGC~612 has been assigned a hybrid radio morphology (both FR~I and FR~II): it has a bright hot spot in the eastern lobe but a diffuse western lobe, and its radio luminosity is near the dividing value between the two types 
(\citealt{2000A&A...363..507G}). 
  
The different locations of the galaxies in the WISE color-color diagram may reflect different evolutionary states. 
While NGC~612 is likely to be at an earlier evolutionary stage than NGC~1316 which has a more evolved galaxy population, it is unclear whether NGC~612 will move upwards first in the 
color-color diagram of Sadler et al. (\citeyear{2014MNRAS.438..796S}; 
towards larger \wonep-\wtwo values more typical of FR~II radio sources hosted by `WISE late-type galaxies') 
or move left (towards FR~I sources hosted by `WISE early-type galaxies').  

\subsection{Stellar masses} \label{sec:SMASS}

The \wise $W1$ (3.4~$\mu$m) band is sensitive to evolved luminous stars that dominate the bolometric luminosity 
of the host galaxy's stellar component. 
Similar to the $K$-band (2.2~$\mu$m), it can be used as a measure of stellar mass 
for nearby ($z < 0.15$) early-type galaxies
(see \citealt{2013AJ....145....6J}).  
Using stellar masses derived from Galaxy And Mass Assembly (GAMA; Taylor et al. \citeyear{2011MNRAS.418.1587T}), \citet{2014ApJ...782...90C} obtained the following relation for resolved low-redshift galaxies: 
\begin{eqnarray}
\log_{10} \left( 
\frac{M_{\rm stellar}}{L_{W1}} 
\right)   = 
- 2.54\,(W_{3.4\,\mu{\rm m}} - W_{4.6\,\mu{\rm m}}) 
- 0.17  \label{eqn5}
\end{eqnarray}
\noindent where 
$L_{W1} = 10^{ -0.4 ({\mathcal M} - {\mathcal M}_{\rm Sun})}$
is the luminosity in the \wone band expressed in solar luminosities ($L_\odot$), 
${\mathcal M}$ is the absolute magnitude of the galaxy in $W1$, 
${\mathcal M}_{\rm Sun} = 3.24$, 
and $(W_{3.4\,\mu{\rm m}} - W_{4.6\,\mu{\rm m}})$ 
reflects the \wise rest-frame color of the galaxy. 
A color correction appropriate to the morphological type and redshift of each galaxy was added 
to the observed $W1-W2$ colors 
(--0.006 mag for NGC~1316 and 
--0.033 mag for NGC~612),  
which were measured within the $1-\sigma$~isophote of the \wtwo images. 
The color correction refers to the cosmic reddening $k$-correction appropriate for each band (\wone and \wtwop, respectively) and was estimated using the spectral energy distribution of an old evolved galaxy for NGC~1316 and an S0 for NGC~612 (both SEDs from the library of \citealt{1998ApJ...509..103S}). 

For NGC~1316, this yields a stellar mass estimate of 
$5.9\times 10^{11}~M_\odot$ for NGC~1316 within 589$''$
(the one-sigma isophotal radius in \wtwop, Table~\ref{tab:ngc1316-sources}). 
From the kinematics of planetary nebulae, \cite{2012A&A...539A..11M} 
constrained the mass distribution in NGC~1316. 
They included a dark matter component, modeled as a pseudo-isothermal distribution. Using their best-fit parameters and after integration of the density profile, we obtain a dark matter mass of $1.92\times 10^{12}$~$M_\odot$, or a stellar to total (stellar + dark matter) mass fraction of 23\% within $589''$.  

For NGC~612, the stellar mass is estimated to be $2.6\times 10^{11}~M_\odot$ 
within $92\farcs6$
(the one-sigma isophotal radius in \wtwop, Table~\ref{tab:ngc612-sources}). 
From observations of optical emission lines in the star-forming disk, 
\cite{1980MNRAS.190P..23G} 
had put an upper limit of $0.8\times10^{12}~M_\odot$ within $46\farcs5$ on the total mass of NGC~612 (after correction to our adopted distance). 
From HI line observations, \cite{2008MNRAS.387..197E} 
could estimate the mass within a larger radius, 115$''$ (70~kpc), to be $2.9\times10^{12}~M_\odot$ (where we assume that the inclination of the galaxy is $90^\circ$).  
Assuming a linear increase of the total mass with radius, this implies that 
the stellar mass represents about 10\% of the total mass in NGC~612. 
The dark matter halo of NGC~612 seems therefore to be more significant than in NGC~1316. 
However, \cite{2012A&A...539A..11M} 
point out the complexity of the kinematics of NGC~1316 with a rise in the rotation curve in the outer parts, which could be a signature of outward transfer of angular momentum produced by the merger. A more sophisticated model such as a non-spherical distribution might be required to obtain a more precise and accurate representation of the mass distribution in this galaxy. 

\subsection{Star formation rates} \label{sec:SFR}

Measurements in the \spitzer MIPS $24~\mu$m band are commonly used to estimate star-formation rates \citep[see][]{2005AAS...207.2109C, 2006ApJ...650..835A, 2007ApJ...666..870C,  2009ApJ...692..556R, 2012ApJ...755..168R}. 
The \wise \wfour (22~$\mu$m) band is similarly sensitive to warm dust continuum from young stars. 
The \wise \wthree (12~$\mu$m) band is sensitive to the 11.3~$\mu$m PAH emission feature, which is well correlated to the star formation rate (SFR) as traced by the extinction-corrected H$\alpha$ luminosity (e.g. \citealt{2016arXiv160101698S}) 
and to dust heated by a combination of young and evolved stars.

We use the relations provided by \citet{2014ApJ...782...90C} 
to estimate the star-formation rates: 

\begin{equation}   
\log_{10}\left( 
\frac{ {\rm SFR} (12\, \mu{\rm m}) }
     { M_\odot~{\rm yr}^{-1} } 
\right) 
= 1.08\, \log_{10}\left(  
\frac{ \nu L_{12\, \mu \rm m}}
     { L_\odot} 
\right)  
- 9.66  
\label{eqn8}
\end{equation}

\begin{equation} 
\log_{10} \left(
\frac{ {\rm SFR} (22\, \mu{\rm m}) }
     { M_\odot~{\rm yr}^{-1} }
\right)
= 0.82\, \log_{10} \left( 
\frac{ \nu L_{22\, \mu \rm m} }
{L_\odot}
\right)
- 7.3 \, .
\label{eqn9}
\end{equation}

\cite{2014ApJ...782...90C} do not provide an estimate of the uncertainties in their scaling relations, but their figures indicate a standard deviation of about 0.3 in the log, which corresponds to a factor of 0.5 to 2 in the absolute value. This is the uncertainty that we attach to the inferred SFRs. 
An additional uncertainty of about 5\% results from the uncertainties in the absolute calibration 
of 4.5\% in $W3$ and 5.7\% in $W4$ \citep{2012wise.rept....1C}, 
multiplied by the factors in the equations above. 
For NGC~1316, we obtain 
$1.7^{+1.7}_{-0.9}~M_{\odot}~{\rm yr}^{-1}$ 
from the 12~$\mu$m measurement and 
$0.77^{+0.77}_{-0.39}~M_{\odot}~{\rm yr}^{-1}$  
from the 22~$\mu$m measurement 
(Table~\ref{tab:ngc1316-sources}). 
For NGC~612, we get 
$19.6^{+19.6}_{-9.8}~M_{\odot}~{\rm yr}^{-1}$   
from the 12~$\mu$m measurement and 
$7.3^{+7.3}_{-3.7}~M_{\odot}~{\rm yr}^{-1}$.  
from the 22~$\mu$m measurement 
(Table~\ref{tab:ngc612-sources}). 
These estimates are consistent with the morphological classifications revealed by the band colors (see Sect.~\ref{sec:band-colors}). 
Because of the
uncertain contribution of heating by evolved stars 
in the 12~$\mu$m band, 
the SFR estimates from the 22~$\mu$m band are likely to be more accurate.  

Far-infrared photometry can be used to obtain an independent estimate of the SFRs. Indeed, the SFR is roughly proportional to the total infrared (TIR) luminosity, ${\rm SFR} = C L_{\rm TIR}$, 
and the constant $C$ lies in the range of $(1.6- 4.1)\times10^{-44}$  
for stellar population models with solar metallicity and constant star formation over a timescale of more than 10~Myr,  
where $L_{\rm TIR}$ is expressed in erg~s$^{-1}$
(\citealt{2013seg..book..419C}; Starburst99, \citealt{1999ApJS..123....3L}). 
Using the TIR luminosities estimated in Sect.~\ref{sectSpitzerIRS}, we derive SFRs 
of 0.30 to 0.77~$M_{\odot}~{\rm yr}^{-1}$ for NGC~1316 
and 
of 7.3 to 18.8~$M_{\odot}~{\rm yr}^{-1}$ for NGC~612. 
Although those various estimates of the SFRs have a large spread, they all indicate that the SFR of NGC~612 is higher of that of NGC~1316 by about one order of magnitude.

\section{Discussion}\label{secDiscussion}

\subsection{Comparison of the two galaxies}

The picture given by the WISE measurements is consistent with NGC~1316 being a massive elliptical galaxy, well represented by a double S\'ersic model, with a low star formation rate and an irregular dust pattern likely due to the recent infall of a small gas-rich companion. 
A two-dimensional fit identifies a central component that may be associated with the AGN; however, 
at the resolution of the WISE images (250 to 500~pc), the complex dust distribution in the central region of NGC~1316 and the AGN cannot be safely separated. 
NGC~612 can be well represented by a bulge and an exponential disk.  The fitting procedure does not identify a central point source in the WISE maps (resolution of 1.6--3.4~kpc). 
The $W3$-to-$W4$ flux density ratio is significantly higher in NGC~1316 ($\sim1.3$) 
than in NGC~612 ($\sim0.8$; 
Tables~\ref{tab:ngc1316-sources} and \ref{tab:ngc612-sources}). 
The starlight from NGC~1316 contributes more strongly to the \wthree emission, 
while in NGC~612 
the higher continuum in \wfour is very likely due to a stronger heating of the dust grains 
by ongoing star formation.  

The core of NGC~612 is about 53 times stronger than that of NGC~1316 (Fornax~A)   
in terms of radio 
power at 4.8~GHz\footnote{
Power of the radio cores at 4.8~GHz: $8.9\times10^{21}$~W~Hz$^{-1}$ for NGC~612 and 
$1.68\times10^{20}$~W~Hz$^{-1}$ for NGC~1316, 
after correction to our adopted distances 
assuming that \cite{1993MNRAS.263.1023M} 
used a distance of 32.7~Mpc for NGC~1316 and of 178~Mpc for NGC~612, 
consistent with the redshifts of the galaxies and their adopted value for the Hubble constant of $H_0 = 50$~km~s$^{-1}$~Mpc$^{-1}$. 
}. 
Those numbers are indications that the nucleus of NGC~612 is currently more active than that of NGC~1316.  
On the other hand, Fornax~A has an inner radio jet, which shows that plasma is still being ejected; no inner jet is clearly seen in NGC~612, although some radio emission is seen in the eastern side (yellow contours in Figs.~\ref{figHSTDSS} and \ref{figdustN612}). 
\cite{1993MNRAS.263.1023M} had pointed out a knotty jet in their lower-resolution observations leading to a hot spot in the eastern lobe. However, deeper radio observations would be required to establish whether the lobes of NGC~612 are still being fed via inner jets.  
From a comparison of X-ray and radio measurements, \cite{1998ApJ...503L..31I} 
argued that the nucleus of NGC~1316 has become dormant during the last 0.1~Gyr, and that the 
galaxy may have the appearance of a normal, inactive elliptical galaxy (with no sign of past nuclear activity) 
in the coming 0.1~Gyr. 
NGC~1316 does not possess much neutral gas that could fuel the AGN 
(about $6\times 10^{8} M_\odot$ of molecular gas and very little HI gas, 
\citealt{2001A&A...376..837H}). 
It does have a small gas-rich companion (NGC~1317, to the North) but this galaxy appears undisturbed and not directly interacting with NGC~1316. 
NGC~612 contains a much larger reservoir of neutral gas (almost 10$^{10}$~$M_\odot$, where about 80\% of it is molecular); 
it also has a neighbor galaxy, NGC~619, located about 400~kpc away with an HI disk that is even larger than its own \citep{2008MNRAS.387..197E} 
and an HI bridge is seen between the two galaxies. 
The near environment of NGC~612 is thus extremely rich in neutral gas that may fuel the nucleus and sustain its activity, especially if 
the two galaxies were to merge in the future and the gas could lose angular momentum. 

\subsection{Challenges to models of galaxy evolution}  

The two galaxies, despite the similarity of their double-lobed radio morphologies, differ in several important ways and raise a number of questions about their formation and future evolution. 
Are they simply at different stages of a common evolution, or do they belong to two distinct source populations that evolve on separate tracks? 

NGC~1316 fits rather well into the standard paradigm of a radio galaxy hosted by an early-type galaxy formed in a merger. 
NGC~612 is clearly more of a challenge to the current models of radio galaxy formation because of its unusual dusty star-forming disk, best revealed in infrared observations. 
Simulations are used to quantify how galaxies assemble their mass, via a combination of smooth accretion and mergers 
(e.g. \citealt{2012A&A...544A..68L}) 
and there is evidence that smooth accretion leads to the formation of disks while mergers have a tendency to destroy disks  
\citep{2015arXiv150205053W}. 
In those simulations, AGN feedback has been identified as a crucial ingredient to prevent massive galaxies from overcooling and growing excessively. Radio-mode feedback in particular affects the extended environment of galaxies by injecting energy via jets and lobes. 
The co-evolution of AGN and their host galaxies is impossible to model exactly because of the enormous range of scales but the nature of the AGN feedback is expected to be different if the surrounding medium is distributed spherically or in a disk \citep{2016AN....337..167W}. 
The dichotomy of FR~I and FR~II radio galaxies observed by \cite{2014MNRAS.438..796S} 
in the WISE color-color diagram seems to indicate that the properties of a radio galaxy are related to those of the host galaxy. It would be interesting to know whether the radio properties are a direct consequence of the morphological types, or whether the radio and optical/mid-IR properties are separate consequences of more fundamental overarching processes such as accretion and mergers in a given environment. 
Are hybrid FR~I/FR~II radio galaxies such as NGC~612 a separate population and how do they form?  

\subsection{Comparison of the two methodologies}
\label{SectCompa}

The two approaches that we have taken to analyze the WISE images are fundamentally heuristic in their nature. 
While it is clear that neither of them will yield a perfect result, they make it possible to gain some insight into the various physical components that make up the observed surface brightness distribution of the galaxies. 
In this section we discuss some of the merits and limitations of each method, 
and comment on potential ways to reduce these limitations and improve both methodologies. 

{\it The fitting method} has the positive and negative aspects of any fitting approach: being model-dependent, it provides quantitative information on the underlying surface brightness distribution; at the same time, the selected models are necessarily a simplification of reality and do not capture the features that they do not include (for instance the warp in NGC~612). The method also requires a convolution with the PSF, which needs to be well determined. Any error in the PSF of the \wone band, which was used to represent the starlight, will affect the fit to the \wone image and the results of the subsequent fits to the other bands, since the geometrical parameters obtained in \wone were kept constant. 
Obviously, errors in the adopted PSF of the other bands may also have an impact on the value of the best-fit amplitude. 
The method, however, does not require any smoothing of the actual images of the galaxies (only the models are convolved with a PSF), whereas the images are smoothed in the scaling method, as discussed below. 

As in any fitting method, there is a danger for the method to converge to a wrong set of parameters, for example if the fit is driven by a ``bad" region of the image. Human intervention can be used to inspect the original images and the residuals and help decide whether some regions should be masked (we did mask some nearby stars). 

Another important issue is the use of a single number, the reduced $\chi^2$, to determine the best-fit parameters. Aside from the fact that the noise may not be truly Gaussian, there might be a covariance between parameters, leading to degenaracies in the fit. 
We have not investigated this. 
A Baysesian inference approach based on Monte Carlo calculations would be more rigorous 
(see, e.g, the new galaxy image decomposition tool GALPHAT, 
\citealt{2011MNRAS.414.1625Y}).  

In the longer-wavelength bands, the fitting method obviously tries to reach the lowest possible residuals across the image, in the chi-square sense.  
In reality, we know that there are features of interest (the dust) that are not included in the model. It would be difficult to model the dust distribution explicitely, especially in a merging galaxy like NGC~1316 where it is known to be very irregular. 
One could, however, devise a more sophisticated method to help the fit converge to a more physical solution, for instance in an interative approach by keeping the positive residuals of the image above a certain threshold, subtracting them from the original \wthree or \wfour images, and re-running GALFIT until the residual image is mostly Gaussian and homogeneous. Alternatively, independent information on the dust could be used as a prior, for example, from optical dust extinction maps.

If the dust emission follows the same spatial distribution as the starlight, the fitting method will not be able to recover it because its contribution would be included in the 
fitted amplitude to \wthree or \wfourp. 
This might be more of an issue for the thermal dust that is likely to be associated with the old stellar population and dominate in \wfourp; in \wthreep, some of the emission is due to PAHs that are excited by young stars. 

Recovering the characteristics of the AGN requires high-resolution images and a high signal-to-noise ratio.  
In our analysis of the WISE images, the fitting method did not lead to the identification a point-like source in the center of NGC~612.  

{\it The scaling method} is more empirical and therefore simpler. 
However, being model-independent, it does not provide the same quantitative insights into the distribution of the starlight. 
On the other hand, using the \wone image rather than a fitted model to it might be a more accurate representation of the starlight. 
To test this, we have applied both methods to the \wtwo image of NGC~1316 and examined the residuals. 
As mentioned earlier, the \wtwo image is similar to the \wone image and is expected to be also dominated by starlight. 
The fitting method yields a residual image that is similar to the one obtained for \wone (right panel of Fig.~\ref{figstarlightN1316}). 
We have also run GALFIT on the \wtwo image without using \wone to constrain the structural parameters. 
Although the reduced chi-square is slightly better and the best-fit parameters slightly different, the residual image presents the same features. In the scaling method, however, the residual image is much more uniform around zero outside the central 30$''$. In the inner region, we see a positive and negative residual suggestive of an offset of about three pixels (about 2$''$) between the smoothed \wone and the \wtwo image. 
When studying the very central regions, additional work to obtain a more precise alignment of the images might lead to improvements.

The scaling method has two obvious sources of uncertainties.  
The first one is the value adopted for the scaling factor, which was taken from the SED of an old E-type galaxy to estimate the stellar contribution. 
By varying the scaling factor and examining the resulting images, we could demonstrate that it is accurate to better than 10\% for the two galaxies considered. 

The other uncertainty comes from smoothing the \wone image to the angular resolution of the \wthree (or \wfour) image. This was done by a simple convolution with a Gaussian of full width half maximum (FWHM) calculated as 
$({\rm FWHM}_{\rm W3}^2 - {\rm FWHM}_{\rm W1}^2)^{1/2}$. 
The PSF in a given band is not exactly equal to the PSF of \wone convolved with this particular Gaussian function.
(The PSF in \wone is more regular than the PSFs in \wthree and \wfourp). 

Finally, the scaling method obviously provides the difference image between the \wthree (or \wfourp) image and the smoothed and scaled \wone image, which can be regarded as an image of the non-stellar component that includes both the AGN and the dust. If the AGN contributes to \wonep, then the flux of the AGN in the final image is reduced by the corresponding amount. 
The scaling method therefore does not separate the dust from the AGN. 
(However, as mentioned earlier, this is difficult to achieve also in the fitting method given the resolution of the WISE images). 

We have also considered an hybrid method, by scaling the best-fit \wone model and subtracting it from \wthree or \wfourp, using the scaling factors used in the scaling method. The resulting images were very similar to the ones obtained in the fitting method. 

Both methods are sensitive to the background subtraction. An inclined plane was included in GALFIT, while in the scaling method the background was assumed to be flat. The inclination of the best-fit plane, however, was negligible. 
 
Finally, the analysis was performed on enhanced-resolution images obtained using the MCM-HiRes method described and tested by \cite{2012AJ....144...68J}. 
Artifacts of the deconvolution and the accuracy of the determined PSFs are limiting factors in how far both methods can be applied. 

\section{Conclusions} \label{sec:conclusion}

We analyzed \wise observations of two nearby radio galaxies, NGC~1316 and NGC~612. Enhanced-resolution images were produced and photometric measurements were performed in the four \wise bands. 
We used GALFIT, a two-dimensional image decomposition software, and a scaling method to separate the stellar from the non-stellar emission and compare the non-stellar component to optical images that show the dust in extinction.  
The main results can be summarized as follows:

\begin{enumerate}

\item The integrated flux density measurements of our enhanced-resolution \wise images of NGC~1316 compare very well with those from the corresponding \spitzer bands in \citet{2005ApJ...633..857D}. 

\item A striking difference in the \spitzer IRS spectra of the central regions of the two galaxies is the strength of the 7.7~$\mu$m PAH feature, which is suppressed relative to the 11.3~$\mu$m PAH feature in NGC~1316, as observed in samples of dusty elliptical galaxies \citep{2008ApJ...684..270K} and low-luminosity AGNs \citep{2007ApJ...656..770S}, 
but is strong in NGC~612. 

\item By testing different models, we found that the superposition of a double S\'{e}rsic is a reasonable representation of the starlight of NGC~1316, which dominates the emission in the $W1$ band. 
For NGC~612, a better fit is obtained by the superposition of a S\'{e}rsic model and exponential disk. However, those models have their limitations, especially in the central regions, but also to represent the disk of NGC~612 that is strongly warped. Therefore, we also used a simple scaling model to separate the stellar from the non-stellar emission of the galaxies in the long-wavelength bands. 
Both methods reveal the morphology of the dust distribution in the galaxies. 
In NGC~1316, the $12~\mu$m emission coincides with the dust patches seen in the HST image in two regions to the NW and SE of the nucleus and observed at 8~$\mu$m with \spitzer \citep{2010ApJ...721.1702L}. 
In NGC~612, the $12~\mu$m and $22~\mu$m emission comes from a warped disk.  

\item The $W3$-to-$W4$ flux density ratio is significantly higher in NGC~1316 than in 
NGC~612. 
This difference may be due to a larger fraction of starlight in the \wthree filter for  
the larger and more luminous galaxy NGC~1316. 
PAH emission at 11.3~$\mu$m may also be stronger in NGC~1316, although this does not seem to be the case in the central region of the galaxies for which \spitzer IRS spectra exist. 
On the other hand, 
the dusty azimuthal arc 
in the NW of the nucleus of NGC~1316 shines brightly in $W3$, which might be due to enhanced PAH line emission. 
To our knowledge, no mid-IR spectra are available from that interesting region 
located just beyond the inner radio jet. The emission in \wfour is mostly due to heated dust;  
it is not surprising that it dominates in NGC~612 that has ongoing star formation thoughout the disk. 

\item 
The two galaxies fall at different locations in the \wise color-color diagram: 
NGC~1316 appears in the region of gas-poor early-type galaxies and FR~I radio sources, whereas
NGC~612 is in the region of more actively star-forming galaxies 
(it is a LIRG) and FR~II sources.  
The star-formation rate of NGC~1316 derived from the measurements in the 22~$\mu$m band is $\sim 0.7\, M_\odot$~yr$^{-1}$, 
while that of NGC~612 is ten times higher. 
NGC~612 has an even higher specific star formation rate (SFR normalized to the stellar mass), 
which indicates that it is still actively forming its stellar population. 

\item  
Those two bright radio-lobed galaxies 
with low-luminosity optical AGN have very different mid-IR properties, which are likely related to their formation processes and evolutionary stages. 
The dust in NGC~1316 is likely due to the recent infall of one or several small gas-rich companions, 
while in NGC~612, gas must have been accreted more continuously to settle in a rotating star-forming disk.  
Radio continuum measurements from the literature \citep{1993MNRAS.263.1023M} 
 indicate that the nucleus of NGC~1316 is 
currently more dormant than that of NGC~612. 
While the nucleus of NGC~1316 may be momentarily revived if the dusty material from the NW and SE clumps makes it to the center, the galaxy is very likely to evolve into a completely passive elliptical within no sign of past activity within a few 0.1~Gyrs, when the synchrotron emission from the 
nucleus and the lobes will have faded away. 
NGC~612, on the other hand, because of its own larger reservoir of molecular and atomic gas and the nearby presence of a gas-rich companion \citep{2008MNRAS.387..197E} 
may grow a larger star-forming disk and/or sustain a level of activity in its nucleus. 
Numerical simulations of the growth of galaxies through accretion and mergers and  
high-resolution observations of dust and gas in radio-loud galaxies will help shed light on 
the evolution of those fascinating sources. 

\end{enumerate} 

\begin{acknowledgements}
We are very grateful to the referee for an exceptionally thorough review and many constructive comments. 
We thank Michelle Cluver, John H. Black and Roy Booth for their interest in this study and their support. 
We also thank Zolt Levay and Paul Goudfrooij for making their HST image of NGC~1316 available to us; 
Raffaella Morganti and Bj\"orn Emonts for sending us their Australia Telescope Compact Array radio map of NGC~612; 
Chien Peng for the help with the GALFIT program; 
Lauranne Lanz for advice on GALFIT and for sending us her FITS image of the high resolution VLA map of NGC~1316. 
The first author acknowledges funding from the SKA (Square Kilometre Array) Africa Postgraduate Bursary Programme and from MIDPREP, an exchange programme between two organizations in Europe (Chalmers University of Technology in Sweden and ASTRON in the Netherlands) and three South African universities (University of Stellenbosch, Rhodes University and University of Cape Town). 
He thanks the Hartebeesthoek Radio Astronomy Observatory, Onsala Space Observatory, and the University of Cape Town for assistance during his regular visits.  
This research has made use of the NASA/IPAC Extragalactic Database (NED) which is operated by the Jet Propulsion Laboratory, California Institute of Technology, under contract with the National Aeronautics and Space Administration.  
\end{acknowledgements}

\bibliographystyle{aa} 
\bibliography{pap}

\end{document}